\documentclass[11pt]{article}
\usepackage[utf8]{inputenc}
\usepackage{caption}
\usepackage{subcaption}
\usepackage[dvipsnames]{xcolor}
\usepackage{hyperref}
\usepackage[T1]{fontenc}
\usepackage{amsmath}
\usepackage{mathtools}
\usepackage{tensor}
\usepackage{braket}
 \usepackage{amsfonts}
\usepackage{setspace} 
\usepackage{amsthm}
\usepackage{graphicx}
\usepackage{xcolor}
\usepackage[section]{placeins}
\usepackage[normalem]{ulem}
\usepackage[margin = 2.5cm]{geometry}
\usepackage{amssymb}
    
\def\nref#1{(\ref{#1})}

\def\mE{\mathbb{E}}

\def\IR{\text{IR}}

\newcommand{\normord}[1]{:\mathrel{#1}:}

\theoremstyle{definition}

\def\be{\begin{equation}}
\def\ee{\end{equation}}
\def\la#1{\label{#1}}

\begin{document}

\thispagestyle{empty}
\begin{center}
    ~\vspace{5mm}

  \vskip 2cm 
  
   {\LARGE \bf 
      Renormalization in Liouville gravity \\ \vspace{3pt} and stochastic inflation
   }

    \vspace{0.5in}

 Jordan Cotler$^1$, Victor Ivo$^2$, and Juan Maldacena$^3$
  
    \vspace{0.5in}
  $^1$
{\it Department of Physics, Harvard University, Cambridge, MA 02138, USA}
   \\
   ~
   \\
  $^2$
{\it  Jadwin Hall, Princeton University,  Princeton, NJ 08540, USA }
   \\
   ~
   \\
  $^3$
{\it   Institute for Advanced Study,  Princeton, NJ 08540, USA }

\end{center}

\vspace{0.5in}

\begin{abstract}

Motivated by inflation and Liouville theory, we consider random $d$-dimensional geometries characterized by an overall scale factor $ds^2 = e^{2 \zeta} dx^2 $ given in terms of a random Gaussian field $\zeta(x)$ with logarithmic correlations. We discuss aspects of the renormalization of the volume element $e^{d \zeta }$, connecting well-known Liouville theory formulas (KPZ) and inflationary ones.   By starting from a fixed physical cutoff and coarse-graining operators to a fixed fiducial cutoff, defined via the flat metric $dx^2$, we provide a direct physical derivation of the KPZ scaling relation.  We point out that the same renormalization problem arises in stochastic inflation, where it is modeled by Brownian motion of the inflaton field.   We show that the breakdown of the renormalization of the Liouville volume when the fluctuation amplitude exceeds a critical value corresponds to the transition to eternal inflation.  In $d = 2$, this matches the familiar $c_m = 1$ barrier in Liouville gravity.   We also discuss connections to mathematical probabilistic approaches to random surfaces.

\end{abstract}
 
\vspace{1in}

\pagebreak

\setcounter{tocdepth}{3}
{\hypersetup{linkcolor=black}\tableofcontents}

\section{Introduction  }
\la{Newintro}

We consider  random geometries   given by  
\be \la{metric}
ds^2 = e^{ 2 \zeta } d {\hat s }^2 
\ee 
where $d{\hat s}^2$ is a fixed $d$-dimensional geometry of Euclidean signature, for example, flat space $d{\hat s}^2 = dx^i dx^i $,  and $\zeta$ is a random Gaussian field with a two-point function of the form
\be \la{ZetaFl}
\langle \zeta(\vec x) \zeta (\vec y ) \rangle = { 2 \over d Q^2 } \log { 1 \over |\vec x - \vec y | } + \cdots 
\ee 
where $Q$ is an arbitrary parameter that sets the strength of the geometric fluctuations. The dots denote possible IR divergences or terms that are subleading at short distances, which will not play a role in our discussion. 

This problem was first considered for $d=2$ in the classic work on two dimensional gravity \cite{Polyakov:1981rd, Knizhnik:1988ak}. It has also been further developed by mathematicians and physicists studying probability theory \cite{Kahane1985, Duplantier:2008prc}, including extensions to higher dimensions \cite{Schiavo:2021unx}, see also \cite{Levy:2018bdc,Kislev:2022}.

\begin{figure}[t!]
    \centering
    \includegraphics[width=0.75\linewidth]{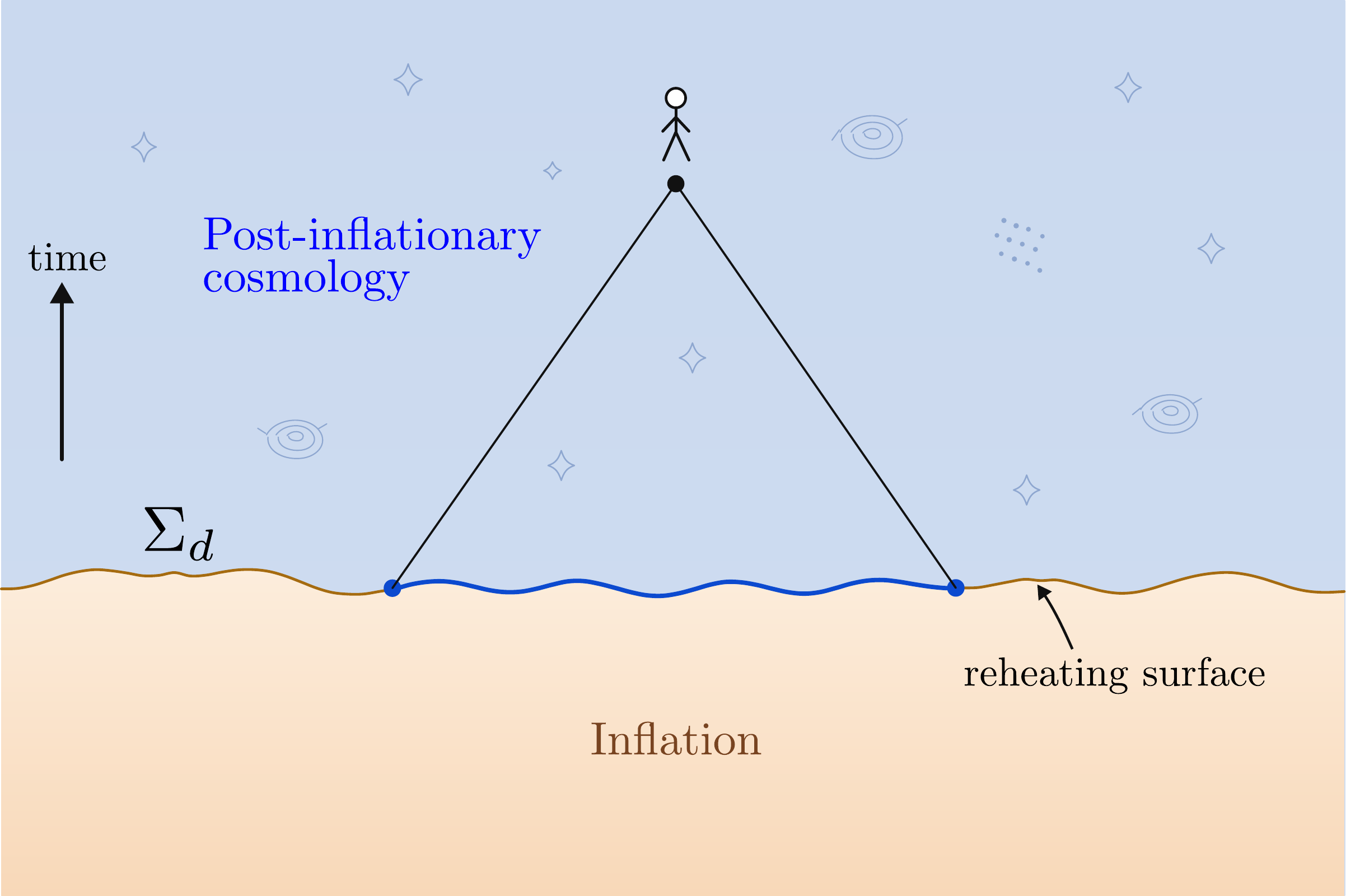}
    \caption{Cartoon of the Penrose diagram for $d+1$ dimensional cosmologies studied in inflation. There is a period of inflation at very early times, which ends at a reheating surface $\Sigma_d$, as in \nref{metric}. The surface $\Sigma_d$  is observable (indirectly) to someone living in the later universe. One role of inflation is to provide a probability measure for the random geometry of $\Sigma_d$. This measure is just a Gaussian measure for the scale factor $\zeta$ in \nref{metric}.   }
    \label{cosmofig}
\end{figure}

Importantly, this problem also arises in a certain limit of the theory of cosmic inflation.  Recall that cosmic inflation produces a three-dimensional geometry on the reheating surface that provides initial conditions for the subsequent hot big bang evolution of the universe; see Figure \ref{cosmofig}. That geometry is random, approximately of the form \nref{metric} with statistical properties approximately set by \nref{ZetaFl}.  Actually, the scale invariant form \nref{ZetaFl} is not exactly correct for the observable scales \cite{Planck:2013jfk}, but it is a good first approximation. Moreover, there is a particular limit of inflation,  a very slow roll limit,  where we can neglect tensor fluctuations and the fluctuations become scale invariant as in \nref{ZetaFl}. This limit was used in \cite{Creminelli:2008es} for studying the transition to slow roll eternal inflation \cite{Linde:1986fd, Starobinsky:1986fx}. 

\begin{figure}[t!]
    \centering
    \includegraphics[width=1\linewidth]{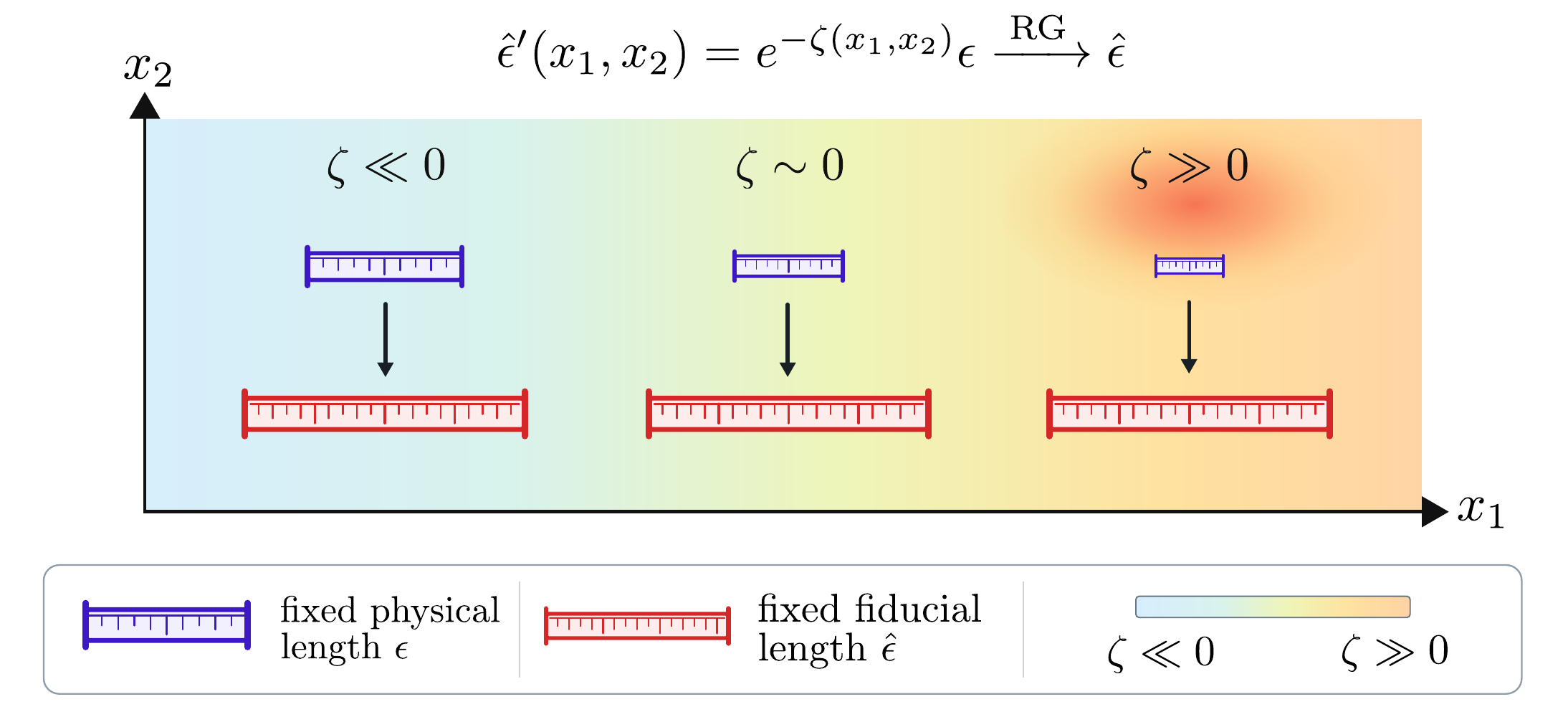}
    \caption{The picture shows an example of the random surface in $d=2$, plotted in fiducial coordinates. The color illustrates the local value of $\zeta$. The purple ruler has a fixed physical size $\epsilon$, so it has a fiducial size $\hat{\epsilon}'$ that varies with $\zeta$. The red ruler has a much larger fixed fiducial size $\hat{\epsilon}$. Our coarse graining can be imagined as integrating out the scales between the two rulers locally.    }
    \label{figrulers}
\end{figure}

To illustrate the goal of our paper, consider the following issue. Take the volume of the surface with metric \nref{metric} to be
\begin{equation}
\label{volnew}
dV=e^{d\zeta}d^{d}x\,.
\end{equation}
At first sight, this is simply the exponential of a Gaussian field defined on the fixed geometry with metric $d\hat s^2$, which we call the fiducial, or ``comoving'', metric. However, as recognized in \cite{Polyakov:1981rd, Knizhnik:1988ak, David:1988hj, Distler:1988jt}, this operator is more complicated than it looks, because one needs to choose a ``ruler'', or a cutoff, in defining the volume. General covariance forces us to pick a cutoff which corresponds to a fixed physical distance $\epsilon$ in the metric $ds^2$ in \nref{metric}. Since physical and fiducial lengths are locally related by $ds=e^\zeta d\hat s$, the same physical ruler has the field-dependent fiducial length 
\begin{equation}
    \hat\epsilon'=e^{-\zeta}\epsilon\,.
    \label{physicalfiducialcutoff}
\end{equation}
This relation is illustrated in Figure \ref{figrulers}. It seems unclear how to work with operators defined with this type of short-distance cutoff, but it was argued in \cite{Polyakov:1981rd, Knizhnik:1988ak, David:1988hj, Distler:1988jt, Levy:2018bdc, Kislev:2022} that the proper expression for the volume involves a substitution of the form 
\be \la{KPZnIntro} 
\exp\!\left(  d \zeta \right) ~~\longrightarrow ~~~ \exp\!\left(  d b Q \zeta_{\hat \epsilon} \right)
 ~~~~~~{\rm with }~~~ b (Q-b) = 1 ~~~{\rm or} ~~~~ 
  b = { Q\over 2} - \sqrt{ {Q^2\over 4 } -1}
\ee 
where the operator on the right-hand side is a conventional exponential operator defined in the fiducial metric, with a constant fiducial cutoff $\hat{\epsilon}$.    

One purpose of our paper is to explain in detail the physical origin of the KPZ formula \nref{KPZnIntro}. More specifically, we will show that it can be understood as
\be \la{KPZOur}
\langle \exp\!\left(  d \zeta_\epsilon \right)  \rangle_{\rm short} \propto \exp\!\left(  d b Q \zeta_{\hat \epsilon} \right) 
\ee 
where the operator $\zeta_{\epsilon}$ on the left-hand side is defined with a small physical regulator $\epsilon$, and $\langle \,\cdot\, \rangle_{\text{short}}$ means that we coarse-grain the operator to obtain a conventional CFT operator, as on the right-hand side, with a fixed fiducial cutoff $\hat{\epsilon}$.   This coarse-graining is an unusual type of renormalization group   transformation, because the total amount of renormalization group time\footnote{Note that we are using the convention that RG time is the logarithmic difference between two scales.} $\tau$ between the two scales
\begin{equation}
\label{RGTi}
e^{\tau}=\frac{\hat{\epsilon}}{\hat{\epsilon}'}=e^{\zeta}\,\frac{\hat{\epsilon}}{\epsilon}
\end{equation}
is $\zeta$ dependent (see Figure \ref{figrulers}). Indeed, an unusual type of $\zeta$-dependent renormalization must be necessary to reproduce the change of $\zeta$ exponents in \nref{KPZnIntro}.

Let us give an intuitive explanation for this exponent change. The random surface \nref{metric} is described by a scale-invariant field $\zeta$, and should therefore be self-similar, e.g.~a fractal. With fractals, the actual volume can depend on the cutoff we use, as in the famous example of measuring the coastline of Britain \cite{Mandelbrot1967}. One way of seeing this for the surface is to take two fixed fiducial scales $\hat{\epsilon}_{2}>\hat{\epsilon}_{1}$. 
Let $\zeta_{\hat \epsilon _{1}}$ and $\zeta_{\hat \epsilon _{2}}$ be regularized fields, which only contain modes with wavelengths bigger than these scales. Then, we find that exponentials of the field obey 
\be  \la{RGFid}
\langle \exp\!\left(  d   \zeta_{\hat \epsilon_1} \right)  \rangle_{\rm short} = \exp \!\left(  d   \tau/Q^2   \right)  \exp \!\left(  d   \zeta_{\hat \epsilon_2} \right) ~,~~~~~~~~{\rm with } ~~~e^{\tau }= { \hat \epsilon_2 \over \hat \epsilon_1} 
\ee 
with $\langle\,\cdot\, \rangle_{\text{short}}$ again standing for integrating all modes between the scales. Notice this is a standard quantum field theory result, since $\zeta$ with a fiducial cutoff is effectively a Gaussian field in a fixed background. It is important for us that the coarse-graining produces a scale dependence via the exponent of the renormalization group time $\tau$. In usual quantum field theory, this dependence can be removed by multiplicative renormalization. The difference in our context is that $\tau$ depends on $\zeta$, and any net $\zeta$ dependence is physical. 

The actual RG between a physical cutoff and a fiducial cutoff is more complicated than \nref{RGFid}. But if $Q$ is large, the result \nref{RGFid} is a good approximation, since the fluctuations of $\zeta$ are small and \nref{RGTi} is approximately constant. Therefore, we can use \nref{RGTi} and \nref{RGFid} to derive
\be \la{kpzapprox}
\langle \exp\!\left( d \zeta_\epsilon \right)  \rangle_{\rm short} \, \propto  \,    \exp \! \left(  d \big(1 + { 1 \over Q^2}\big) \zeta_{\hat \epsilon }  \right), ~~~~ Q\gg1
\ee 
which agrees at leading order with \nref{KPZnIntro} once we use $bQ = 1+ b^2 = 1 + 1/Q^2 +\cdots $. Equation \nref{kpzapprox} is just saying that regions where the coarse-grained field is larger have more Gaussian fluctuations, which makes their volume even larger; see Figure \ref{figrulers}.

A goal of this paper is to explain how we can perform this subtle renormalization group transformation at finite values of $Q$ as well. The technical difficulty is that the amount of renormalization group time $\tau$ fluctuates with the value of the short-distance fluctuations in $\zeta$. We will show how to proceed by relating this coarse-graining to an average over first-passage random walks, using ideas introduced in \cite{Creminelli:2008es, Dubovsky:2008rf, Duplantier:2008prc, Gwynne:2019bhi}. In fact, we will be able to derive the more general KPZ formula for other exponentials $e^{ d h \zeta }$.  

Note that $b$  in \nref{KPZnIntro} becomes complex when $Q<2$. In $d=2$ Liouville theory, this is beyond the so-called $c_m=1$ barrier, where the terminology arises from the fact that in two-dimensional gravity coupled to matter with central charge $c_m$ we have $6Q^2 = 25-c_m$ \cite{Distler:1988jt, David:1988hj, Seiberg:1990eb}. 
In this regime, we will find that the integrating-out procedure actually diverges, in which case the physical volume with a fixed fiducial regulator is not complex, but infinite. Therefore, we see a drastic difference between renormalization in ordinary quantum field theory and the renormalization of the same operator from the perspective of quantum gravity. In quantum gravity, it seems that the volume becomes infinite, even for a finite physical cutoff!  Closely related behavior was previously found in the context of stochastic inflation \cite{Creminelli:2008es,Dubovsky:2008rf}, to which we now turn. 

One of our observations is that  \nref{KPZnIntro}   is identical to the formula derived in \cite{Dubovsky:2008rf} for the growth of the volume of an inflationary region, obtained by integrating out quantum fluctuations. In that context, the computation was guided by the Starobinsky picture of inflation, in which the scalar field in a Hubble region follows a stochastic path driven by both classical evolution and quantum fluctuations \cite{Starobinsky:1986fx}. We can equivalently interpret this stochastic evolution as a renormalization procedure for the probability distribution for the scale factor of the metric on the final slice. This is the usual statement that time during inflation is encoded in terms of distance scales in the final probability distribution. In other words, starting with the scale factor at the reheating surface, we can smooth it over larger and larger regions in order to get a picture of inflaton fluctuations at earlier and earlier times. 

Using the fractal interpretation of the random surfaces in \nref{metric}, we can view the factor of $dbQ$ as an effective fractal dimension characterizing the scaling of the coarse-grained volume (and we will also discuss a more refined notion of fractal dimension later). For the reheating surface of our universe, we do not have three dimensions of space but rather 
\be \la{DimUni}
\tilde d = d bQ =  3+(9.47 \pm 0.10)\times 10^{-9}
\ee 
dimensions. Here, we used \nref{zetDimF} (see \cite{Dubovsky:2008rf}) and the value of the scalar fluctuation amplitude for our universe \cite{Planck:2013jfk}, which is too small to give an appreciable correction.  We view the fractal correction as an interesting curiosity. 

The value $Q=2$, which marks
the barrier in Liouville theory, is also precisely the
transition point to the eternal inflation regime derived in \cite{Creminelli:2008es, Dubovsky:2008rf}. Indeed, inflation implies that the reheating surface volume is infinite, which is precisely what we obtained from the random surface theory side. This is interesting because it relates two seemingly different singular problems, namely Liouville beyond the $c_{m}=1$ barrier (and its higher-dimensional analogs) and eternal inflation. Therefore, perhaps by learning about one, we can obtain clues about the other.  For example, in \cite{Creminelli:2008es} it was argued that in the eternal inflation regime $Q<2$, there is a positive probability that the volume is infinite. This feature is clear from the inflation side \cite{Creminelli:2008es}, but less clear from the random surface theory side, although there is a discussion reminiscent of eternal inflation in \cite{Gwynne:2019bhi} that we review later.

Our paper is organized as follows: In section \ref{newrenliouv},  we discuss the physical picture behind this renormalization in detail from the purely $d$ dimensional point of view, based on equations \nref{metric} and \nref{ZetaFl}. In section \ref{sec:inflation}, we discuss how the previous procedure is reflected in the inflationary computation of \cite{Creminelli:2008es, Dubovsky:2008rf}. In section \ref{sec:mathlit}, we discuss some connections to the mathematical literature.

\section{Renormalization in Liouville (or Logarithmic) gravity}
\label{newrenliouv}

In this section, we discuss a physical derivation of the KPZ formula \nref{KPZ2}. Of course, this formula has been derived many times. Our only point, inspired by \cite{Creminelli:2008es, Dubovsky:2008rf, Duplantier:2008prc, Gwynne:2019bhi}, is to explain in physical terms the origin of the formula. The derivation we present is longer than the short one in \cite{David:1988hj, Distler:1988jt}, but we think it is worthwhile to spell it out because it provides additional insight.

\subsection{Generalities }

We are interested in a gravity theory where the metric takes the form 
\be \la{metricgen}
ds^2 = e^{ 2 \zeta } d \hat s^2 
\ee 
where $d\hat s^2$ is a fixed reference metric, which we call the ``fiducial'' metric.  When we say that we have a ``gravity theory'', we mean a few things.  First,  we are integrating over metrics; in this case, we restrict the integral to conformally flat metrics, so that the integral is just over the scale factor parametrized by $\zeta$. Second, we impose a short-distance cutoff using the physical metric $ds^2$. Namely, for a physical cutoff $\epsilon$, the physical distances measured with $ds^2$ should be larger than $\epsilon$. Finally, we should treat reparametrization as a gauge symmetry. After choosing conformal coordinates and fixing the fiducial metric, the residual diffeomorphisms are conformal reparameterizations, which are then quotiented. 

The description of the theory has the following redundancy 
\begin{equation}
\label{weylsym}
\zeta \rightarrow \zeta-\omega\,,~~~~ \hat{\gamma}_{ab} \rightarrow e^{2\omega}\hat{\gamma}_{ab}\,,
\end{equation}
which leaves the physical metric $ds^2 = e^{2 \zeta} d\hat{s}^2$ unchanged.  Physical quantities should therefore be independent of the choice of fiducial representative $\hat \gamma_{ab}$.  In a gauge-fixed description, the action and measure for $\zeta$ may each depend on the fiducial metric $\hat \gamma_{ab}$ and need not be separately invariant under the above transformation. However, this fiducial metric dependence must cancel in physical quantities.  In two dimensions, this is the familiar statement that the total Weyl anomaly vanishes, $c_{\rm total} = 0$.  Note that, since $\hat \gamma_{ab}$ is a fixed fiducial object rather than an integration variable, \nref{weylsym} does not represent an additional gauge redundancy whose volume should be divided out.

All of the features we discussed are essentially kinematic properties, independent of the particular Lagrangian that we consider. They are independent of the particular statistics or measure we have for $\zeta$.  For the standard two-dimensional Liouville theory, we have a nice and local Lagrangian for $\zeta$. However, for the $d=3$ inflationary case that we want to consider, that is not the case. What is important is that physical quantities depend only on the metric $ds^2$, and not on the arbitrary decomposition into $\zeta$ and $\hat\gamma_{ab}$.  Thus the final statistical description must transform consistently under \nref{weylsym}.

After these initial remarks, we restrict to a flat fiducial metric from now on,
\be \la{FlatFid}
d{\hat s}^2 = dx^i dx^i 
\ee 
which is without loss of generality, since we will only need to use the short-distance behavior of $\zeta$. We will not consider the most general action. We only consider theories  in which the resulting measure for $\zeta$ is Gaussian, with a two-point function of the form 
\be \la{ZetaFl2}
\langle \zeta(\vec x ) \zeta (\vec y) \rangle = { 2 \over d Q^2 } \log { 1 \over | \vec x - \vec y | } + \cdots 
\ee 
at short distances.

This makes the field $\zeta$ a ``logarithmic field'' in two related ways. First, its correlators are logarithmic. Second, the transformation law  in 
\nref{weylsym} makes it into what is called a ``logarithmic operator'' \cite{Gurarie1993,Zamolodchikov2004}. 

Note that the transformation law \nref{weylsym} determines the normalization of the operator $\zeta$. We will fix this normalization once and for all. This then uniquely defines the parameter $Q$ (or $Q^2$) that appears in \nref{ZetaFl2}.\footnote{It is common to normalize the Liouville field as $\phi = Q \zeta$. In that normalization, the analog of \nref{weylsym} becomes $Q$ dependent and that can become the definition of $Q$.}${}^{,}$\footnote{In the context of Liouville theory, \cite{David:1988hj, Distler:1988jt} discussed a renormalization of the action due to the change of metric used in the definition of the path integral. We are defining $Q$ here via \nref{ZetaFl2}, which is the final correlator, including all such effects. We will not discuss the path integral directly but rather simply assume that, after all such effects are included, the renormalized measure gives the correlator \nref{ZetaFl2}.
}

As is familiar in two dimensional Liouville theory, we can instead consider the path integral over $\zeta$ with the fiducial metric held fixed.  This gives the corresponding fixed-background conformal field theory. This is different from the gravity theory because the cutoffs are defined using the flat metric $d\hat s^2$. However, there can be a relatively simple relation between the two problems \cite{David:1988hj, Distler:1988jt, Seiberg:1990eb}, which we will review below.  In \nref{ZetaFl2}, the two-point function is written in fiducial coordinates. The fact that these are subject to conformal gauge transformations is related to the dots in \nref{ZetaFl2}. We will not attempt to be more definite about them because only the short-distance part of the correlators is important for our purposes. A more extensive discussion will be found in \cite{Second}.    
  
In Liouville theory we normally also consider a ``cosmological constant term'' in the action, which is an integral over the volume. As we mention below, this term is not important for determining the KPZ formula, so we will ignore it in our discussion. Furthermore, in the inflationary discussion of section \ref{sec:inflation}, this Liouville cosmological  constant  term is naturally absent  from the $d$-dimensional statistical theory of $\zeta$; it should not be confused with the effective $(d+1)$-dimensional cosmological constant that drives the inflationary expansion.  Perhaps, instead of calling this ``Liouville'' gravity we should call it ``Logarithmic'' gravity, because only \nref{ZetaFl2} is relevant, which also implies that $\zeta$ itself is what is normally called a ``logarithmic'' field \cite{Gurarie1993,Zamolodchikov2004}.

\subsection{The traditional argument for the KPZ formula}

Here, we review the traditional argument for the KPZ formula. The actual argument given in \cite{David:1988hj, Distler:1988jt, Seiberg:1990eb,Levy:2018bdc,Kislev:2022} is slightly different because it involves appealing to a matter theory. However, it is conceptually the same. 

The starting point is an operator of the form 
\be  \la{VHop}
e^{ d h \zeta }
\ee 
in the full quantum gravity theory. This is an operator which, under the transformation \nref{weylsym}, has scaling dimension 
\be \la{DimOph} 
\Delta =  d \,h\,.
\ee

Now, we can imagine that we integrate out all the short-distance modes up to a comoving or fiducial scale $\hat \epsilon $. Then we expect to get a more conventional operator in that flat space CFT.  The idea in \cite{David:1988hj, Distler:1988jt} is simply to determine the correct operator by demanding that it has the right dimension, namely \nref{DimOph} under the conformal rescaling 
\be  \la{zetaCh}
\zeta \to \zeta - \lambda ~,~~~~~~~ x \to e^{\lambda} x\,.
\ee 
Making an ansatz
\be \la{AlCa}
\normord{\,e^{ d \alpha_{h} Q \zeta }}
 \ee 
we find that its dimension is 
\be 
{ \Delta \over d } =   Q \alpha_{h}  - \alpha_{h}^2   
\ee 
where the first term comes from the shift of $\zeta $ in \nref{zetaCh} and the second term comes from the anomalous scaling of the normal-ordering subtraction. Equating this scaling dimension with \nref{DimOph} we get the KPZ relation 
\be \la{KPZ2}
h = \alpha_{h} ( Q - \alpha_{h} ) ~,~~~~~~{\rm or } ~~~~ \alpha_h = { Q \over 2} - \sqrt{ {Q^2 \over 4 } - h }\,.
\ee

As a comment, note that usually, the KPZ relation is discussed by considering a combined operator of total dimension $d$ which can be integrated as  
 \be 
 \int d^d x \left[   e^{ d  \alpha_{h} Q \zeta } O_{\Delta_m}(x)  \right]_{\rm fid} ~,~~~~~~~ d = d \alpha_h ( Q \  - \alpha_{h}  ) + \Delta_m 
  \ee
  where both are defined with fiducial cutoffs; the $\zeta$-dependent prefactor is the gravitational dressing of the matter operator. However, we are now also saying that in the original gravity theory with a physical cutoff, we should think of the total operator as being 
  \be \la{IntMat}
  \int d^d x \left[   e^{ d  h \zeta } \right] _{\rm phys}  \left[ O_{\Delta_m}(x)\right] _{\rm fid}  = \int d^d x \left[   e^{ d  \alpha_h Q \zeta }  O_{\Delta_m}(x)\right] _{\rm fid} 
    \ee 
    with $h$ given by \nref{KPZ2}. In the physical cutoff description, the corresponding integrated operator has   $dh + \Delta_m = d $. Note that we have included all the $\zeta$ dependence in the first factor, and the operator $O_{\Delta_{m}}(x)$ should be viewed as one renormalized with the fiducial metric.  

For the volume operator $h = 1$, and we conventionally denote $\alpha_{h = 1} = b$.  Thus we have
\be 
1 =b (Q-b) ~,~~~~~~~~~b = { Q \over 2} - \sqrt{ { Q^2 \over 4 } -1 }\,.
\ee 
Note that our perspective is that $Q$ is the fundamental parameter, since it sets the amplitude of the correctly normalized field $\zeta$ for the metric fluctuations, while $b$ is a derived parameter. In some of the Liouville theory literature, the formulas are written taking $b$ as the fundamental parameter. This is perfectly fine if we think about the Liouville CFT, but it is a bit funny from the Liouville gravity point of view.

Note that we talk about Liouville gravity here but the essential feature that determined this formula was the short distance form of the correlator \nref{ZetaFl2}. In particular, the Liouville potential was not important, for the following reason.  The Liouville potential is given by the integral $\int d^d x \, e^{ d b Q \zeta }$ and it is not important when $\zeta \to -\infty $. So we can imagine determining the dimension \nref{AlCa} in that region of the field space. In other words, we can view the problem of determining a primary operator $e^{ d \alpha Q  \zeta}$ as a scattering problem, where the region $\zeta \to -\infty $ is an asymptotic region, where it is easy to determine the dimension, see \cite{Ginsparg:1993is}.  Of course, if one is interested in other computations in Liouville gravity, then the potential will be important. The only point we are emphasizing here is that it is not important for determining the dimension of exponential operators. Similarly, the renormalization argument we will present later in section \ref{renarg} applies in the region where the potential is not important.

Though this is certainly a reasonable derivation of the KPZ formula \nref{KPZ2}, one can still be confused by the fact that the volume operator went from $e^{d \zeta} $ to $e^{ d b Q \zeta }$. What is the physical origin of this different scaling? Notice that this means that when we make the surface bigger by taking $\zeta \to \zeta + \gamma$, the coarse-grained volume does not scale by $e^{ d \gamma } $ but by $e^{ d \gamma b Q }$. It is therefore natural to define an  ``$\zeta$-scaling''   fractal dimension  
\be \la{zeDim}
\tilde d =  d b Q = d(  1 + b^2 )\,.
 \ee 
And as $Q $ goes from $\infty $ to $Q=2$, $b$ goes from zero to 1. This means that the volume scales as if the surface has dimension $\tilde d $, which ranges from $d$ to $2d$, a fact observed in \cite{Dubovsky:2008rf} for a related inflation problem, as we will review later. Equation \nref{zeDim} is the general formula underlying the estimate in \nref{DimUni}. There, we specialized to $d=3$ and used the observed amplitude of scalar fluctuations in our universe \cite{Planck:2013jfk} to determine the corresponding value of $Q$.

For the volume operator, $\alpha_{h = 1} = b$ becomes complex for $Q < 2$.  More generally, for $h > 0$, we see that $\alpha_h$ becomes complex when $Q < 2\sqrt{h}$. Why does this happen? What is the physical interpretation?

In part to answer these questions and in part to connect with cosmology, we will discuss the renormalization process in more detail to hopefully clarify the physical origin of this formula.

\subsection{Detailed renormalization argument }
\label{renarg}

In this section, we will discuss in detail the renormalization of the operator $e^{d h \zeta}$.  One should take our argument as a reasonable physics derivation, not a rigorous math result.   The argument in this section can be understood as the renormalization group interpretation of the stochastic inflation problem considered in  \cite{Creminelli:2008es, Dubovsky:2008rf}, and we will discuss the connection between the two in a later section. It is also related to \cite{Duplantier:2008prc}; see sections \nref{dyadsqr} and \nref{fracdim}.

We consider the operator   
\begin{equation}
\label{phyreg}
O_{h,\text{phy}}=e^{d h\zeta_{L}}f(L)
\end{equation}
where $\zeta_L$ is regulated at a fixed physical distance scale, i.e.
\begin{equation} \la{PhysCut}
\zeta_{L} \text{ has fixed }\textbf{physical}\text{ distance cutoff } \epsilon_{\text{phy}}=e^{L}
\end{equation} 
By ``physical cutoff'', we mean that $\epsilon_{\text{phy}}$ corresponds to a fixed length scale as measured by the random surface metric $ds^{2}$
\begin{equation}\la{PhysCut2}
ds^2 = e^{2 \zeta } d{\hat s}^2 ~,~~~d{\hat s}^2 =dx^idx^i
\end{equation}
which is related to a fixed fiducial metric $d\hat{s}^{2}$ via \nref{PhysCut2}. The function $f(L)$ is a UV subtraction that will be determined later. Due to the fluctuations of $\zeta $ \nref{ZetaFl2}, a fixed physical cutoff corresponds to a fluctuating cutoff for the fiducial $x$ coordinates in \nref{PhysCut}. In other words, defining the fiducial UV cutoff as $\hat \epsilon_{\rm UV} \equiv e^{ -\tau_{\rm UV} } $, we have 
\begin{equation}
\label{endpt1}
\epsilon_{\text{phy}}=e^{L}= \hat \epsilon_{\rm UV} e^{\zeta_{L}}
=e^{-\tau_{\text{UV}}}e^{\zeta_{L}} 
\quad \Longrightarrow \quad \zeta_{L}-L=\tau_{\text{UV}}
\end{equation}

Our goal is to transform the operator $e^{ d h \zeta} $ defined with the physical cutoff into a more conventional CFT operator defined in the fiducial coordinates, which has a $\zeta$-independent cutoff which is defined purely in terms of the fiducial coordinates. 

\begin{figure}[t!]
    \centering
    \includegraphics[width=\linewidth]{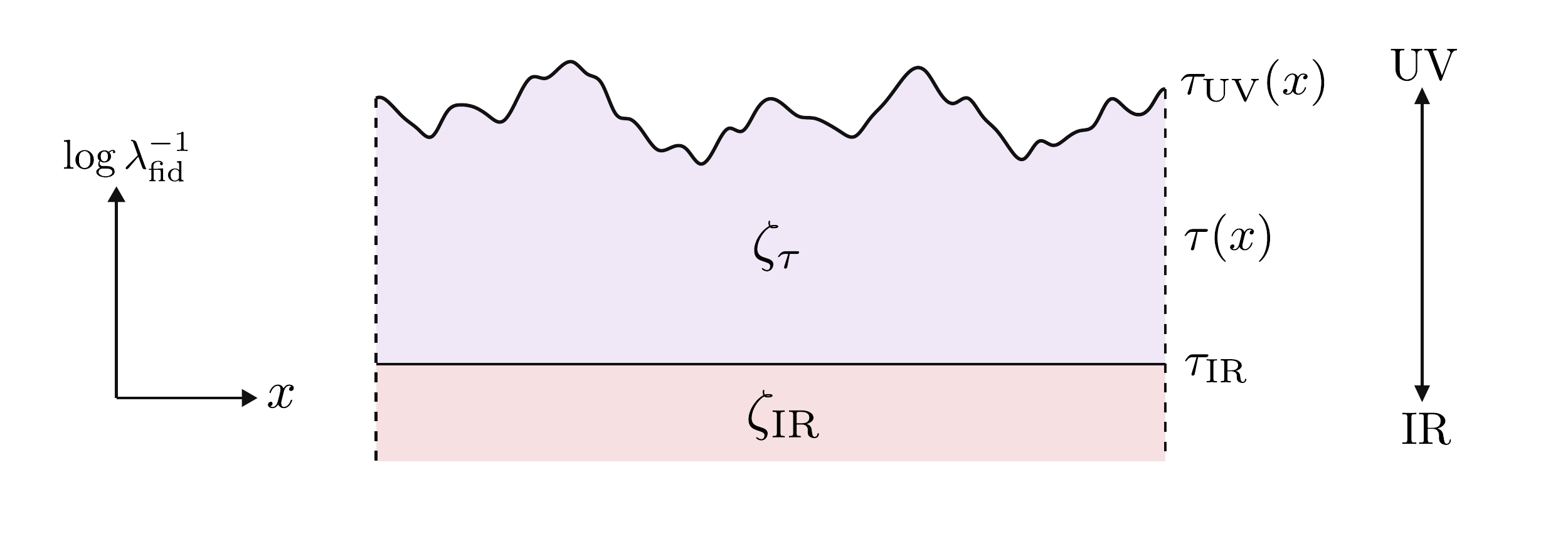}
    \caption{The field is split according to fiducial wavelength. The vertical axis is the log of the inverse fiducial wavelength of various modes. The modes in red, which have fiducial wavelengths larger than $e^{-\tau_{\text{IR}}}$, are in the coarse-grained field $\zeta_{\text{IR}}$. Shorter wavelengths, drawn in purple, are in $\zeta_{\tau}$, up until a UV fiducial wavelength $e^{-\tau_{\text{UV}}}$, which is fixed locally by the condition that its physical length is $e^{L}$. Here $\tau(x)=\tau_{\text{UV}}(x)-\tau_{\text{IR}}$ is the local amount of RG time between the two scales.}
    \label{rgpicture}
\end{figure}

We will achieve this by performing a renormalization group flow between the UV scale and an IR scale, where we integrate out modes until we reach the IR cutoff $\hat \epsilon_{\rm IR} \equiv e^{ -\tau_{\rm IR}}$ that is fixed in the fiducial metric and therefore independent of $\zeta$; see Figure~\ref{rgpicture}. At this IR scale, we will obtain a conventional CFT operator.

For this purpose, we split the regulated field $\zeta_{L}$ into two parts. The first is a coarse-grained field $\zeta_{\text{IR}}$, where
\begin{equation}
\zeta_{\text{IR}} \text{ has fixed }\textbf{fiducial}\text{ distance cutoff }
\hat \epsilon_{\rm IR} = e^{ - \tau_{\rm IR}}
\end{equation}
plus a remainder $\zeta_{\tau}$ containing shorter modes as
\begin{equation}
\label{zetasplit}
\zeta_{L}=\zeta_{\text{IR}}+\zeta_{\tau}\,.
\end{equation}
We assume that the total RG flow time is large enough that we can define this split. In other words, 
\be 
\tau \equiv  \tau_{\rm UV} - \tau_{\rm IR} 
\ee 
is almost always positive, even after we take into account the fluctuations of  $\tau_{\rm UV} $, see Figure \ref{rgpicture}.   
 It is useful to define $N_0$ as 
\begin{equation} \la{DefNz}
e^{N_{0}} \equiv \frac{ \hat \epsilon_{\rm IR}e^{\zeta_{\text{IR}}}}{\epsilon_{\text{phy}}}  ~~~~~{\rm or}~~~~N_{0}=\zeta_{\text{IR}}-\tau_{\rm {IR}}-L\,.
\end{equation}
Here $N_0$ is the amount of RG time one would infer from the coarse-grained field $\zeta_{\rm IR}$ alone, before accounting for the short-distance fluctuations. We assume
\be \la{NzLarge}
  N_{0} \gg 1
\ee 
so that there is parametrically a large range of scales to integrate out.
Note that $N_{0}$ is not yet the actual RG time between the two scales.  That time also depends on $\zeta_\tau$ and is given by 
\be \label{brownendpt}
\tau = \tau_{\rm UV}- \tau_{\rm IR} = \text{log}\bigg(\frac{\hat \epsilon_{\rm IR}}{\hat \epsilon_{\rm UV} }\bigg) = N_0 + \zeta_\tau 
\ee 
and we want $\tau$ to be typically large.  Equation \nref{brownendpt} also shows that coarse-graining over $\zeta_{\tau}$ should be highly non-trivial, since we are integrating over a field $\zeta_{\tau}$ whose number of modes depends on the field itself.  We now explain how to do this.  

First, we are assuming that, at short distances, $\zeta$ is effectively Gaussian and log-correlated. This implies that we can think of $\zeta_\tau$ as a random variable which is a function of $\tau$ and which is executing a Brownian motion in $\tau$, at least for large $\tau$. The intuition for the Brownian motion is that we can think of the field $\zeta_{\tau}$ as a sum of independent Gaussian modes at various scales. For large $\tau$, the correlation of nearby scales, induced by a choice of filter\footnote{For an example of deviations from pure Brownian motion induced by a filter choice, see \cite{bertacco2025liouville}.} to define $\zeta_{L}$, and the possible discreteness of the modes should become negligible. 

In this regime, we can thus approximate $\zeta_{\tau}$ as a sum of infinitesimal uncorrelated Gaussian shells, which is the defining feature of Brownian motion. One can fix the variance of each shell by noting that, under rescaling the coordinates as $x'=e^{-\tau }x$ (see \cite{Chatterjee:2024phq}), the covariance of $\zeta$ at short distances in \nref{ZetaFl} changes as
\begin{equation}
\label{resczeta}
\langle \zeta\big(e^{-\tau}x\big)\zeta(e^{-\tau}y)\rangle=\frac{2}{dQ^{2}}\log \frac{e^{\tau}}{|x-y|}+ \cdots =\langle \zeta(x)\zeta(y)\rangle+\frac{2\tau }{dQ^{2}}+ \cdots
\end{equation}
One can think of the short-distance two-point function of $\zeta$ as the sum over the variance of modes that cannot resolve the two points. The difference between the two sides of \nref{resczeta} thus comes from the variance of modes in $\tau$ log-scales that resolve $|x-y|$ but not $e^{-\tau}|x-y|$. In our approximation, this therefore fixes the variance of short-wavelength modes of $\zeta$ in a log-scale shell $\delta \tau$ as
\begin{equation}
\langle \zeta_{\delta \tau}(x)^{2}\rangle=\frac{2\delta \tau}{dQ^{2}}\,.
\end{equation}

\begin{figure}[t!]
    \centering
    \includegraphics[width=0.8\linewidth]{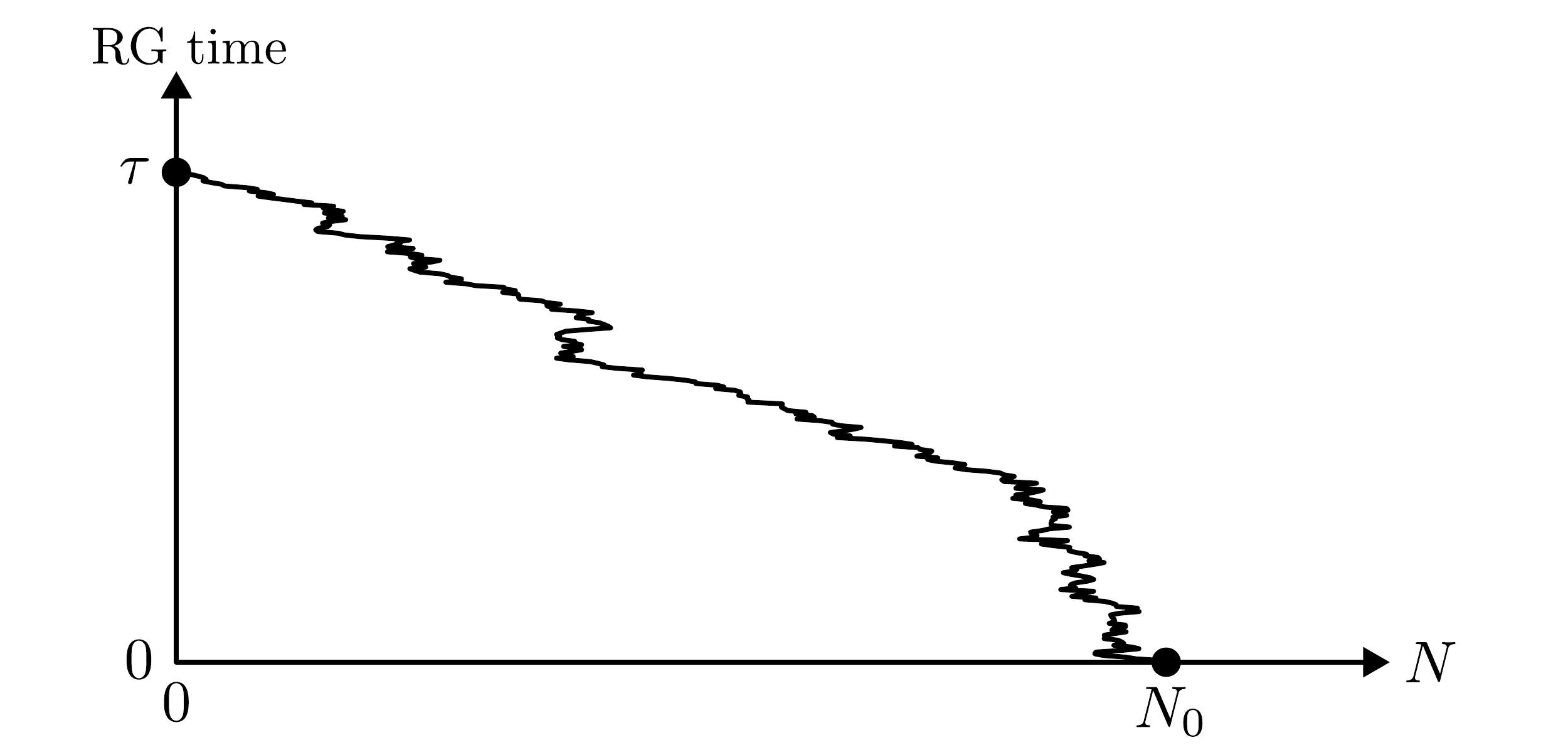}
    \caption{The dynamics of the random motion described by \nref{rdmotion}. The horizontal axis is the variable $N$, and the vertical one is time. At $\tau=0$, the motion starts at $N=N_{0}$, and it evolves by \nref{rdmotion} until it hits $N=0$, where the motion stops.}
    \label{randommotionfig}
\end{figure}

Using these observations, we can view equation \nref{brownendpt} as follows. We define a variable $N$ as
\begin{equation}
N(\tau)\equiv N_{0}-\tau+\zeta_{\tau}
\end{equation}
which starts at $N_{0}$ at $\tau=0$  and then stochastically evolves as $\tau$ progresses, undergoing a net downward drift with Brownian noise. Each time step corresponds to adding a few extra modes to the field, resulting in the change
\begin{equation}
\label{rdmotion}
\delta N=-\delta \tau+\zeta_{\delta \tau}\,,~~\text{ with } \langle \zeta_{\delta \tau}\rangle=0\,,~~~~ \langle \zeta_{\delta \tau}^{2}\rangle=\frac{2\delta \tau}{dQ^{2}}
\end{equation}
with $\zeta_{\delta \tau}$ acting as a white noise term.  The motion stops when we reach $N=0$, which corresponds to the equation \nref{brownendpt}; see Figure \ref{randommotionfig}. Moreover, it is physically reasonable to impose that it should stop at the first $\tau$ such that the equation holds.  The reason is that continuing the motion beyond $N=0$ would generally introduce modes below the physical cutoff. Note that the stopping time $\tau$ is a random variable. 

Therefore, equation \nref{brownendpt} defines a first passage problem at $N=0$ for the random motion described by \nref{rdmotion}, which started from $N_{0}=\zeta_{\text{IR}}-L-\tau_{\rm {IR}}$ at $\tau=0$. Thus, coarse-graining over the fast modes $\zeta_{\tau}$ should be equivalent to averaging over the random paths \nref{rdmotion} in this first-passage problem.

With this in mind, we now integrate out the degrees of freedom of the exponential operator \nref{phyreg} defined with a physical cutoff, in order to obtain an operator with a fixed fiducial cutoff $e^{-\tau_{\IR}}$. The idea is to use the endpoint equation to rewrite \nref{brownendpt} in terms of $\tau$ and the fixed cutoff variables $L$ and $\tau_{\IR}$. After using \nref{DefNz} and \nref{brownendpt},  the resulting operator with a fixed fiducial cutoff becomes
\begin{equation}
\label{prefidreg}
O_{h,\text{fid}}=\langle O_{h,\text{phy}}\rangle_{\text{fast modes}}=e^{d h(L+\tau_{\text{IR}})}f(L)\mE[e^{d h\tau}]
\end{equation}
with $\mE$ denoting an average over the random paths satisfying the first passage problem.

Therefore, the only calculation left to do is to compute the first passage average of $e^{d h\tau}$ over the random paths we discussed. This was done in \cite{Creminelli:2008es}, and we review it in Appendix \ref{statrdmotion}. The conclusion is that the probability distribution for the stopping time $\tau $ over these paths is given by
\begin{equation}
\label{pbrown}
p(\tau)=N_{0}\sqrt{\frac{dQ^{2}}{4\pi \tau^{3}}}\,e^{-\frac{dQ^{2}(\tau-N_{0})^{2}}{4\tau}}
~,~~~~~~~~ \int_0^\infty d\tau\, p(\tau) = 1\,.
\end{equation}

As expected, for large $Q$  the paths are peaked around $\tau = N_0$, which is the naive amount of  RG time between the two scales when we neglect the fluctuations in $\zeta_\tau$.  For $\tau \gg N_0$, the stopping time distribution has the asymptotic form
\be 
p(\tau) \sim \tau^{-{\frac{3}{2}}}e^{-\frac{dQ^{2}\tau}{4}} ~,~~~~~{\rm for } ~~\tau \gg N_0 \,.
\ee  
For $h > 0$, the moment $\mE[e^{d h \tau}]$ is therefore convergent precisely when $Q \geq 2\sqrt{h}$. For $h \leq 0$, there is no corresponding large-$\tau$ divergence. When the moment converges, inserting \nref{pbrown} in  \nref{prefidreg} we find that (see Appendix \ref{statrdmotion}) 
\begin{equation}
\label{expdeltaa}
\mE[e^{d h\tau}]=\int_{0}^{\infty} d\tau\, p(\tau)e^{d h\tau}=e^{d\alpha_{h} Q(\zeta_{\text{IR}}-L-\tau_{\text{IR}})}
\end{equation}
where $\alpha_{h}$ is the smallest root of
\begin{equation} \la{KPZhVa}
h=\alpha_{h}(Q-\alpha_{h}) ~,~~~~~~~{\rm or }~~~~\alpha_{h}=\frac{Q}{2}-\sqrt{\frac{Q^{2}}{4}-h}\,.
\end{equation}
Note that this is the same $\alpha_{h}$ that appears in the usual KPZ relation \nref{KPZ2} of Liouville theory. The calculation also automatically picks the correct root for $\alpha_{h}$, which is the smallest one. Outside the region where the integral is convergent, the expression for $\alpha_h$ becomes complex.  This does not mean that the underlying positive random variable becomes complex; rather, the closed-form expression is then an analytic continuation of a moment that is now divergent.  The square-root branch point of $\alpha_h$ thus coincides with the loss of convergence of the corresponding first-passage moment.

Therefore, at this point we have already confirmed that the $\zeta$ dependence of the operator \nref{prefidreg} is the same as that of the usual normal-ordered operator $\normord{\,e^{d\alpha_{h} Q \zeta}}$, defined in the fiducial cutoff theory. However, we also want to check that the final coarse-grained operator has the correct subtraction, expected from normal ordering. Before we do so, note that the final answer should not depend on the arbitrary physical cutoff scale $L$, so one should pick $f(L)$ in \nref{prefidreg} in order to cancel the $L$ dependence of the expression. We do so by picking
\begin{equation}
f(L)=e^{d\alpha_{h}^{2}L}
\end{equation}
which fixes the final form of $O_{h,\text{fid}}$ to be
\begin{equation}
O_{h,\text{fid}}=e^{d\alpha_{h} Q\zeta_{\text{IR}}}e^{-d\alpha_{h}^{2}\tau_{\text{IR}}}\,.
\end{equation}

The last fact we need is what the expected normal-ordering subtraction should be for $e^{d\alpha_{h} Q \zeta_{\text{IR}}}$. Since $\zeta_{\text{IR}}$ has a fiducial cutoff equal to $e^{-\tau_{\text{IR}}}$, its regulated coincident two-point function should be equal to the two-point function at separation equal to that scale, that is
\begin{equation}
\label{Gfid}
G_{\hat \epsilon}=\langle \zeta_{\text{IR}}^{2}(x)\rangle\approx \frac{2}{dQ^{2}}\log \frac{1}{e^{-\tau_{\text{IR}}}}=\frac{2\tau_{\text{IR}}}{dQ^{2}}\,.
\end{equation}

The conclusion is then that our final operator is
\begin{equation} \la{FinRel}
\langle e^{d h  \zeta } |_ {\text{phy}}\rangle_{\text{short}}=O_{h,\text{fid}}=e^{d\alpha_{h} Q \zeta_{\text{IR}}}e^{-d\alpha_{h}^{2}\tau_{\text{IR}}}=e^{d\alpha_{h} Q \zeta_{\text{IR}}}e^{-\frac{d^{2}\alpha_{h}^{2}Q^{2}}{2}G_{\hat \epsilon}}=\,\normord{\,e^{d\alpha_{h} Q \zeta}}
\end{equation}
as we wanted to argue. In the first line, we plugged in the result from \nref{expdeltaa} and the chosen value for $f(L)$. Then, we used the relation between $\tau_{\text{IR}}$ and the coincident two-point function of $\zeta_{\text{IR}}$ to show that this factor of $\tau_{\text{IR}}$ is precisely the correct one to subtract the two-point function divergence of $\langle e^{d\alpha_{h} Q\zeta_{\text{IR}}}\rangle\sim e^{\frac{1}{2}d^{2}\alpha_{h}^{2}Q^{2}G_{\hat \epsilon}}$.

Note that for $h=1$, equation \nref{FinRel} implies that the volume element of the metric \nref{metric}, defined with the natural physical cutoff, becomes an operator in the fiducial description with a non-trivial renormalized exponent.
\be 
  \la{FinVol}
\langle e^{d  \zeta } |_ {\text{phy}}\rangle_{\text{short}}= \,\normord{\,e^{db Q \zeta}} ~,~~~~~~~b = { Q \over 2 } - \sqrt{ { Q^2 \over 4 } -1 }\,.
\end{equation}
For $d=2$, the right-hand side is the usual cosmological constant operator of Liouville theory for the given value of $Q$, which is expected from its interpretation as the volume of a random surface. 

However, one interesting aspect of this calculation is that it makes precise what breaks down for the volume operator in Liouville theory when $Q<2$. In this regime, the moment $\mE[e^{d\tau}]$ in \nref{expdeltaa} diverges, and hence the relation \nref{FinRel} cannot be obtained by this coarse-graining procedure.\footnote{In this paper, we will not discuss the borderline case $Q=2$, or equivalently $b=1$. We will restrict attention to the regimes $Q>2$ and $Q<2$.} The fact that $b$ becomes complex is a symptom of analytically continuing the formal expression beyond its domain of convergence. The underlying volume for an individual realization has not become complex; instead, the expectation value required to define the corresponding operator at fixed fiducial cutoff diverges. Notably, this divergence occurs even though the original theory is defined with a finite physical UV cutoff.  The origin of this behavior is the gravitational nature of the problem: since the relation between physical and fiducial scales depends on the fluctuating field $\zeta$, coarse-graining to a fixed fiducial cutoff requires averaging over a random, and potentially arbitrarily large, range of RG scales.  This mechanism has no direct analog in the usual renormalization of an operator in QFT on a fixed background, where the relation between UV and IR cutoffs is fixed independently of the quantum fields. Thus what fails is our attempt to represent the physically regulated theory by an ordinary operator in the fixed fiducial background description. It remains possible that a formulation which does not pass through a fixed fiducial background continues to make sense.

Let us stress that, from the point of view of random geometry, we are not guaranteed that we will have the emergence of a smooth fiducial space. This is a special feature of the $Q>2$ system; the average over short distance modes produces a smooth fiducial space. This is the case when we have a large value of $\tau $ in the renormalization group transformation, so that we are studying features of the surface at distances much longer than the physical UV cutoff. This is in contrast with quantum field theory, where the structure of space is given as a starting point. In that case, we are guaranteed to be able to go to longer distances.

Of course, once we have established \nref{FinRel} \nref{FinVol} for a single operator,  
\be \la{Multiple}
\langle e^{ d h \zeta(x_1) }\cdots e^{ d h \zeta(x_n) }|_{\rm phy} \rangle_{\rm short} =  \,:e^{ d \alpha_h Q\zeta(x_1) }: \cdots :e^{ d \alpha_h Q \zeta(x_n) }:  
\ee 
as long as the separations are much larger than the fiducial cutoff, $|x_i-x_j|\gg \hat\epsilon$. For such separations, the UV modes being integrated out are local to each insertion, so there are no additional shared short-distance divergences coupling distinct insertions. The local renormalizations therefore factorize.

\subsection{Further comments about the renormalization}
\label{simpcom}

In this section, we give more intuition as to why the physically regulated operator $e^{d h \zeta_L}|_{\rm phy}$ becomes $e^{d \alpha_h Q \zeta_{\rm IR}}$
when written with a fixed fiducial cutoff. In particular, we would like to understand the physical origin of why $\alpha_{h}$ is a non-analytic function of the variance of $\zeta$, set by $Q^{-2}$.
 
Let us give a qualitative argument that is valid for large $N_0$. Consider the fast mode $\zeta_\tau$, which for fixed $\tau$ is  Gaussian with variance 
 \be \la{GaussZet}
 \langle \zeta_\tau \zeta_\tau\rangle   = { 2 \over d Q^2 } \,\tau ~,~~~~~{\rm or } ~~~~p(\zeta_\tau) \propto \exp\left( - { d Q^2 \over 4 } { \zeta_\tau^2  \over \tau } \right).
 \ee
 In our problem, however, $\tau$ is also a variable, and it is constrained by 
 $\zeta_\tau = \tau  -N_0$. Therefore, we can represent
 the expectation value for the volume as 
\begin{equation}
\la{ProbNG}
\begin{aligned}
\langle e^{d h \tau} \rangle &\propto \int d\tau  \int d \zeta_\tau  \,\delta( N_0 + \zeta_\tau - \tau ) \exp\!\left( - { d Q^2 \over 4 } { \zeta_\tau^2  \over \tau } \right) \exp\!\left( d h \tau \right)\\
&=\int d \zeta_\tau \exp\!\left( - { d Q^2 \over 4 } { \zeta_\tau^2  \over (\zeta_\tau+N_0) } \right) \exp\!\left( d h (\zeta_{\tau}+N_{0}) \right)\,.
\end{aligned}
\end{equation}
 This expression captures the relevant behavior of the exact integral in \nref{expdeltaa}; we have only neglected prefactors and the limits of integration. After integrating over $\tau$, equation \nref{ProbNG} becomes an ordinary integral over $\zeta_\tau$. Note that once we impose the relation between $\tau$ and $\zeta_{\tau}$, the distribution for $\zeta_{\tau}$ is no longer Gaussian. The intuition for this non-Gaussianity is that  a longer trajectory $\tau$ allows a larger fluctuation to accumulate, while a larger fluctuation in turn makes the trajectory last longer. At large $N_{0}$, we can study this interplay by  evaluating \nref{ProbNG} with a saddle point approximation. The saddle point equation for $\zeta_{\tau}$ is 
\begin{equation}
-\frac{d\,Q^{2}\zeta_{\tau}}{2(\zeta_{\tau}+N_{0})}+\frac{d\,Q^{2}\zeta_{\tau}^{2}}{4(\zeta_{\tau}+N_{0})^{2}}+d \,h=\frac{d\,Q^{2}}{4}\bigg(1-\frac{\zeta_{\tau}}{\zeta_{\tau}+N_{0}}\bigg)^{2}+d \,h-\frac{d\,Q^{2}}{4}=0
\end{equation}
which leads to
\begin{equation}
\label{saddlept}\zeta_{\tau}\big|_{\text{saddle}}=N_{0}\bigg(\frac{1}{\sqrt{1-\frac{4 h}{Q^{2}}}}-1\bigg)=\frac{2\alpha_{h}}{Q-2\alpha_{h}}N_{0}=\frac{2\alpha_{h}}{Q}\tau\big|_{\text{saddle}}
\end{equation}
which is non-analytic in $Q$, due to the non-trivial way that $\zeta_{\tau}$ enters \nref{ProbNG}. 
 As $Q\to2\sqrt h$ from above, both
$\tau_{\rm saddle}$ and $\zeta_{\tau,\rm saddle}$ run to infinity. Namely,   the 
coarse-graining becomes dominated by arbitrarily long RG
trajectories. 

Another interesting point is that the noise $\zeta_\tau$ at the saddle \nref{saddlept} grows linearly with $N_{0}$, and thus with the saddle time $\tau$. This implies that the saddle $\zeta_{\tau}$ is very atypical at large $N_{0}$. More specifically, using \nref{GaussZet}, its probability decays exponentially with $N_{0}$ as 
\begin{equation}
p|_{\text{saddle}}\sim e^{-d\alpha_{h}^{2}\tau|_{\text{saddle}}}=e^{-\frac{d\alpha_{h}^{2}Q}{Q-2\alpha_{h}}N_{0}}\,.
\end{equation}
What drives $\zeta_\tau$ into this range is the exponential weighting by $e^{dh\zeta_\tau}$.  Thus at large $N_0$ the weighted coarse-graining is dominated by rare upward fluctuations with $\zeta_\tau = O(N_0)$ instead of by typical fluctuations of size $O(\sqrt{N_0})$.  Because $N_0$ depends linearly on $\zeta_{\rm IR}$, these rare configurations modify the coefficient of $\zeta_{\rm IR}$ in the effective exponential $e^{d \alpha_h Q \zeta_{\rm IR}}$.  More specifically, the final $N_0$ dependence results from the competition between this exponential enhancement and the exponentially small probability of such long trajectories.

This is connected with the following fact, reviewed in more detail in section \ref{sec:mathlit}. The continuum limit of measures for these  random surfaces are supported at special points with large upward fluctuations called ``thick points''. Thick points are very unlikely configurations, where the field configuration diverges as the log of the cutoff when the cutoff goes to zero. Nevertheless, they contribute to the volume because the volume is exponentially enhanced, compensating for the unlikelihood of these events.

\section{Inflation as a logarithmic conformal field theory}
\la{sec:inflation}

In this section, we describe a limit of slow roll inflation where we find a statistical theory of geometries with a random scale factor. 
In this case, the renormalization group time evolution described in section \ref{newrenliouv} is very closely related to actual Lorentzian time evolution. We will also explain how the volume renormalization of \cite{Dubovsky:2008rf} is related to the KPZ scaling relation. 

\subsection{Generalities about inflation }
\label{geninfl}

We have in mind the following cosmological history. We have an early period of slow-roll inflation followed by a standard hot big bang evolution, which we will not describe in detail; see Figure \ref{cosmofig}. The inflationary period is driven by an inflaton field $\phi$. To be more specific, the theory effectively describing the universe in this period is gravity with a minimally coupled scalar $\phi$ with a Lorentzian action
\begin{equation}
I_{L}=\frac{M_{\rm pl}^{d-1}}{2}\int d^{d+1} x \sqrt{g}\,R+\int d^{d+1} x \sqrt{g}\left[-\frac{1}{2}(\nabla \phi)^{2}-V(\phi)\right].
\end{equation}
Classically, we can write the metric as
\begin{equation}
ds^{2}|_{\text{spacetime}}=-dt^{2}+a(t)^{2}dx^{2}\,.
\end{equation}
The $x$ coordinates are sometimes called ``comoving'' coordinates. 

During the slow-roll regime of inflation, the Friedmann equations can be approximated as  
\begin{equation}
\dot{\phi} \approx -\frac{\partial_{\phi}V}{dH(\phi)}\,,~~~~ H^{2}\approx \frac{2V}{d(d-1)M_{\rm pl}^{d-1}}\,.
\end{equation}
When the inflaton approaches a certain critical value, these approximations cease to be valid,  inflation ends, and the universe is supposed to enter the radiation-dominated hot big bang phase. For simplicity, we assume that the reheating is instantaneous. In this approximation, there is a spacelike surface $\Sigma_d$ where inflation ends. The surface $\Sigma_d$ is called the ``reheating surface''. We treat the FLRW cosmology in the future of $\Sigma_d$ as a classical measuring apparatus that measures its geometry. This just means that the observer in Figure \ref{cosmofig} can determine the shape of this surface through appropriate cosmological observations. 

To be more specific,  classically $\Sigma_d$ is a fixed surface of very small curvature. For most purposes, we can treat it as a subregion of flat space. However, due to quantum fluctuations during inflation, $\Sigma_d$ has fluctuations in its shape. We can divide them into two types: scalar fluctuations $\zeta$ and tensor fluctuations $h_{ij}$. 

In spatially flat gauge, the scalar fluctuations are encoded in fluctuations of the inflaton field $\phi$. Transforming to the uniform-$\phi$ reheating surface converts these into fluctuations $\zeta$ of the local scale factor. The tensor fluctuations come from gravitons produced during inflation. 

In \cite{Creminelli:2008es}, the authors introduce an interesting theoretical limit of inflation, which we call ``linear roll'', where tensor fluctuations are absent, and the scalar fluctuations are scale invariant. In this limit, one takes
\begin{equation} \la{LinRol}
\textbf{Linear Roll limit}:~~M_{\rm pl}\rightarrow \infty~\text{ and }~H,~\partial_{\phi}V = \text{constant}
\end{equation}
This can also be viewed as a very slow roll limit, where the slow roll parameters are much smaller than what is strictly necessary for ordinary slow roll inflation.  This limit puts us in a regime that can approach slow-roll eternal inflation \cite{Linde:1986fd, Starobinsky:1986fx}. In fact, that was the motivation in \cite{Creminelli:2008es}. 

\begin{figure}[t!]
    \centering
    \includegraphics[width=0.9\linewidth]{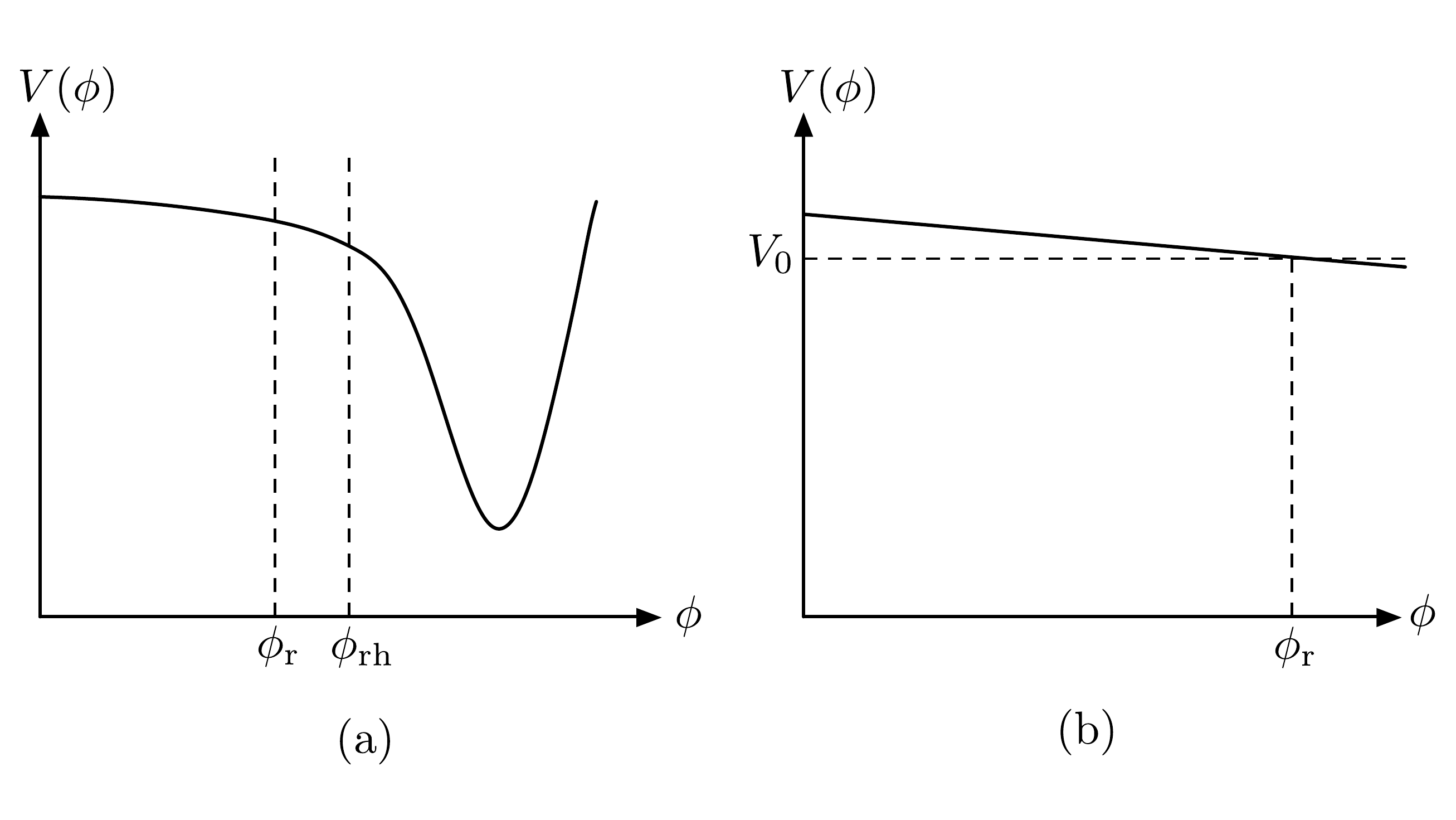}
    \caption{(a) A potential $V(\phi)$ for the inflaton in slow-roll inflation. The slow-roll regime continues up until $\phi_{\text{rh}}$, where reheating happens. (b) The inflaton potential in approximately the linear roll regime. One assumes that $V(\phi)$ has a large offset and a comparatively small constant slope. We assume reheating happens at $\phi_{r}$, which one can imagine is slightly before the actual reheating, at $\phi_{\text{rh}}$, in a more realistic potential. This way, one can glue ``linear roll'' evolution to a faster roll evolution plus the necessary reheating dynamics later.}
    \label{inflpotfig}
\end{figure}

This limit can be achieved by taking the following inflaton potential (see Figure \ref{inflpotfig})  
\begin{equation} \la{LinPot}
V(\phi)=V_0 + (\phi-\phi_r)V' ~,~~~~{\rm with }
~~~ V_0 \equiv \frac{d(d-1)}{2}M_{\rm pl}^{d-1}H^{2} \end{equation}
with $V'$ a fixed number and taking $M_{\rm pl}\rightarrow \infty$. The background spacetime in this limit is exact de Sitter space. There are no gravitons, and thus no tensor fluctuations, because they go to zero at infinite $M_{\rm pl}$. However, the scalar fluctuations sourced by the inflaton remain finite, since their amplitude is controlled by $H$ and $\dot{\phi}$, which are held fixed in this limit.

In the linear roll limit \nref{LinRol}, the  spacetime  metric is exactly de-Sitter 
\begin{equation}
\label{spctmet}
ds^{2}= H^{-2} \left[ - d \tau^2 + e^{ 2 \tau } dx^2 \right].
\end{equation}
Notice that $\tau = Ht$ measures time in units of e-folds.

Due to the absence of tensor fluctuations, the metric on the reheating surface can be written as  
\be \la{metricCos}
ds^2 = e^{ 2 \zeta} dx^2 
\ee 
where $\zeta$ is a random field. We have absorbed a uniform constant into $\zeta$. Moreover, since the potential is linear and the metric is exactly de Sitter, the power spectrum of $\zeta$ is perfectly scale invariant, which means that it scales in Fourier space as $\langle \zeta_{k}\zeta_{-k}\rangle\sim k^{-d}$. This implies that the position space two-point function of $\zeta$ has a log singularity as $x \rightarrow y$. Using the explicit two-point function for $\zeta$, one can fix the prefactor and conclude that
\begin{equation}
\label{2ptlinroll}
\langle \zeta(\vec x)\zeta(\vec y)\rangle=\frac{2}{dQ^{2}}\log \frac{1}{|\vec x-\vec y|}+ \cdots ~~,~\text{ with }~Q^{2}=\frac{\dot{\phi}^{2}}{H^{d+1}}\text{Vol}(S^{d+1})
\end{equation}
where the dots denote terms that are less relevant at short distances (and include an irrelevant IR divergence).\footnote{One might wonder whether there are order one corrections to the expression for $Q^2$ that come from one loop corrections to the wavefunction of the universe. We will argue in \cite{Second} that there are no such corrections.}

The relation between the inflaton and $\zeta$  at scales larger than the horizon is given by 
\be \la{ZetPhiRel}
\zeta(\vec x ) =  \left. \tau - { H\over \dot \phi }  \left[ \phi(x,\tau) - \phi_r \right]  \right|_{\rm super ~horizon}     ~,~~~~~~ \tau = H t\,.
\ee 
Note that the right-hand side is constant at super horizon scales.   This relation is clear at the end of inflation where $\phi = \phi_r$. The fact that $\phi$ evolves linearly in time, for a linear potential, ensures that it is also valid no matter what time we evaluate it, as long as we consider superhorizon scales so that we can neglect the time evolution of the decaying modes. Of course, which scales are superhorizon scales depends on time, and a more thorough discussion of this dependence is important for the discussion below. 

Note that though the fluctuations have a quantum origin during inflation, we can view the final reheating geometry $\Sigma_d$ \nref{metricCos} as a random geometry generated by a log-correlated field of the type discussed previously. In other words, $\zeta$ is a classical random variable.  In particular, the full discussion of the renormalization of the volume and other exponentials in section \ref{newrenliouv} also applies here.

As a side comment, note that this linear roll limit has several properties in common with the nearly dS$_2$ \cite{Maldacena:2019cbz,Cotler:2019nbi} (and also AdS$_2$ \cite{Maldacena:2016upp, Engelsoy:2016xyb, Jensen:2016pah}) limit that is described by JT-gravity. Here we have exact de-Sitter and a scalar field that breaks that exact symmetry and determines the location of the reheating surface. In nearly-dS$_2$ (or AdS$_2$) we have a scalar field that breaks the symmetries of the dS$_2$ space and determines the location of the reheating surface (or the AdS$_2$ boundary). The analog of the dS$_2$ reparametrization modes is here  $ \zeta(x) $, though in inflation, for odd $d$,  we cannot obtain it from a purely local action on the boundary. Interestingly, the description of JT gravity with a boundary at a finite location discussed in \cite{Ferrari:2024kpz} involves a one dimensional version ($d=1$) of the problem discussed in this paper.

\subsection{Starobinsky stochastic evolution as field-dependent renormalization}
\label{stocrg}

In \cite{Creminelli:2008es}, the authors developed a method to find the precise slope of the potential that separates ordinary slow roll inflation from eternal inflation. In the process, they perform an analysis that is very closely related to what we discussed in section \ref{newrenliouv}.  

The authors of \cite{Creminelli:2008es}  introduce a time-dependent coarse-graining for the inflaton field 
\begin{equation}  \la{PhiReg}
\phi_{L}(x,\tau)=\int d^{d} y\,e^{d(\tau - L)}\,f(e^{\tau - L} |x-y|)\phi(y,\tau)\,,\qquad \int d^d z \,f(z) = 1\,,
\end{equation}
where $\phi$ is the uncoarse-grained inflaton field. We can choose any smooth filter function $f(z)$ localized around $|z| \lesssim 1$. Thus $\phi_L(x,\tau)$ averages the inflaton over a comoving region of linear size $|x-y| \sim e^{L-\tau}$, or equivalently a physical region of linear size $e^L H^{-1}$. Since a Hubble radius corresponds to the comoving scale $e^{-\tau}$, this region contains parametrically $e^{dL}$ Hubble volumes. The parameter $L$ is fixed throughout the stochastic evolution. More precisely, one takes $L$ sufficiently large so that the modes entering the coarse-grained field are well outside the horizon \cite{Creminelli:2008es}. Since the final renormalized quantities of interest are independent of this fixed choice, we will suppress $L$ in what follows and, for notational convenience, set $L=0$. This should be understood simply as a choice of origin for the logarithmic coarse-graining scale.

The authors in \cite{Creminelli:2008es} imagine that they know the value of this coarse-grained inflaton at some initial time $\tau_{\rm 0}  $.   
Fixing the value of the inflaton at $\tau=\tau_{0}$ is thus the same as fixing all modes of the inflaton which have a physical wavelength bigger than a few horizon sizes at that time. The regulated field \nref{PhiReg} does not have any shorter modes at this time. Since time is constant in the slice, this is equivalent to putting a comoving or fiducial cutoff on the coordinates $x$,  
\be 
\hat \epsilon_{\rm IR} =   e^{ -\tau_{0} } = e^{ - \tau_{\rm IR}} ~,~~~~~\tau_0 = \tau_{\rm IR}\,.
\ee 
This is the scale of the horizon at that time.

\begin{figure}[t!]
    \centering
    \includegraphics[width=0.8\linewidth]{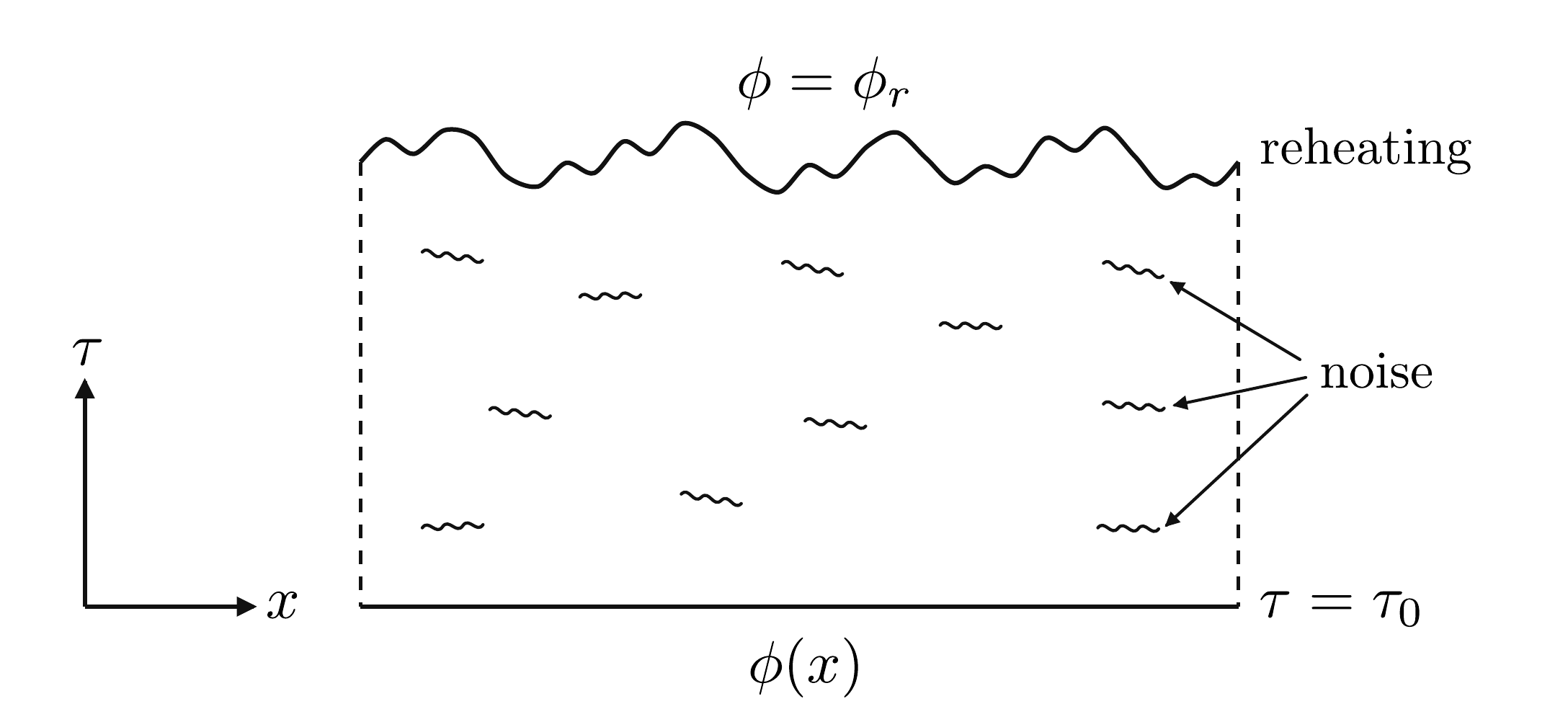}
    \caption{First passage random motion in stochastic inflation. A profile for the inflaton is fixed at $\tau=\tau_{0}$, and it starts evolving by classical evolution and from a series of random kicks from modes entering the cutoff at later times. These can be viewed as arising from quantum fluctuations that become classical, or are ``measured'', as they exit the horizon. The motion stops locally once the inflaton achieves reheating at $\phi=\phi_{r}$. The time until reheating is random, due to the random noise from modes entering the cutoff. }
    \label{stocinffig}
\end{figure}

We are interested in following the evolution of the coarse-grained field at a given point $x$. As time evolves and the universe expands, shorter comoving wavelengths cross the coarse-graining scale and begin to contribute to \nref{PhiReg}. Since these modes are new quantum variables, their values were not fixed by specifying the coarse-grained field on the initial slice, so in the effective long-wavelength description we average over them. Successive modes therefore appear as stochastic kicks to the coarse-grained inflaton.
Following \cite{Starobinsky:1986fx}, this evolution is described by a stochastic process
\begin{equation} \la{StarEv}
\delta \phi=\dot{\phi}\,\delta t+\delta W \,,~~~~~~~ \dot \phi = - { V' \over d H } = {\rm constant}
\end{equation}
with $\delta W$ a white noise term fixed by the two-point function of $\phi$. $\delta W$ is a Gaussian random variable with
\begin{equation} \la{NoiseW}
\langle \delta W\rangle=0\,,~~\langle (\delta W)^{2}\rangle=\frac{2H^{d}}{d\,\text{Vol}(S^{d+1})}\,\delta t\,.
\end{equation}

At each point, this stochastic evolution ends when $\phi(x,\tau)$ first reaches the reheating value $\phi_{r}$, as depicted in Figure \ref{stocinffig}. The corresponding reheating time $\tau_{\rm UV}$ is therefore a stochastic variable, and the evolution defines a first passage problem for the random motion of the inflaton \cite{Creminelli:2008es}, similar to the one we discussed in section \ref{newrenliouv}. To see how the problems are essentially identical, we can define the variable $N$
\begin{equation} \la{NdefPhi}
N(\phi)=\frac{H(\phi_{r}-\phi)}{\dot{\phi}}
\end{equation}
which is the classical amount of e-folds required to go from $\phi$ to $\phi_r$, if we were to ignore the extra fluctuations described by \nref{NoiseW}. Thus, reheating happens at $N=0$. Using \nref{NdefPhi}, we can then rewrite \nref{StarEv} as
\begin{equation}\la{NwithEta}
\delta N=-\delta \tau+\delta \eta
\end{equation}
where the white noise $\delta \eta$ satisfies
\begin{equation}
\langle \delta \eta(x,t)\rangle=0\,,~~~~\langle(\delta \eta(x,t))^{2}\rangle=\frac{2}{dQ^{2}}\delta \tau\,,~~~\text{ with } Q^{2}=\frac{\dot{\phi}^{2}}{H^{d+1}}\,\text{Vol}(S^{d+1})\,.
\end{equation}

Note that $\delta \eta$ is the same as the variable  $\zeta_{\delta \tau}$ we had in section \ref{newrenliouv}. This is because a small fluctuation $\delta \phi $ translates into a small fluctuation $ \delta \zeta   = - { H \delta \phi  / \dot \phi } $, see \nref{ZetPhiRel}. This makes manifest that this stochastic motion is exactly the one described in section \ref{newrenliouv}, where the ``time variable'' was log scales. 

The dictionary with section \ref{newrenliouv} can now be made explicit. The coarse-grained inflaton profile at the initial time $\tau_0=\tau_{\rm IR}$ determines the long-wavelength field $\zeta_{\rm IR}$ through \nref{ZetPhiRel}. This is the quantity that is parametrized in \cite{Creminelli:2008es, Dubovsky:2008rf} by the ``initial value of the inflaton''. The modes that subsequently cross the stochastic coarse-graining scale are precisely the short-distance modes $\zeta_\tau$ that were integrated out in section \ref{newrenliouv}. Equivalently, the stochastic calculation averages over these later modes in order to express quantities such as the reheating volume purely in terms of the initial, coarse-grained configuration $\zeta_{\rm IR}$.

The local reheating time $\tau_{\rm UV}(x)$ determines the upper endpoint of this evolution, so that the (stochastic) duration $\tau_{\rm UV}(x)-\tau_{\rm IR}$ is precisely the random RG time of the previous section. At reheating, $\phi=\phi_r$, and \nref{ZetPhiRel} gives $\zeta(x)=\tau_{\rm UV}(x)$, up to the uniform constant that we have already absorbed into $\zeta$. The resulting field is what we denoted by $\zeta_L$ in section \ref{newrenliouv}: the full field regulated at a fixed physical short-distance scale. In the stochastic description, this physical regulator is set by the fixed coarse-graining scale at reheating, of order $e^L H^{-1}$, with the fixed parameter $L$ suppressed in our notation as described above. At an earlier time $\tau$, by contrast, the stochastic description gives a coarse-grained version of $\zeta$ containing only modes with comoving wavelengths larger than the corresponding scale $\Delta x \sim e^{-\tau}$.

One important point is that what is called the ``initial value of the inflaton'' in \cite{Dubovsky:2008rf} is really an observable property of the late universe. An observer in the late universe, as in Figure \ref{cosmofig}, can in principle reconstruct the geometry of the reheating surface, choose suitable comoving coordinates, and coarse-grain $\zeta$ over a comoving scale $\hat\epsilon_{\rm IR}$, thereby retaining only wavelengths larger than $\hat\epsilon_{\rm IR}$.  This defines precisely the field $\zeta_{\rm IR}$ used above. Then the actual formula for the volume of the reheating surface, if she could measure it, or wanted to compute it, would be given by the integral of $e^{ d b Q \zeta_{\rm IR} } $, with $b$ as in \nref{FinVol}, and with $Q$ in \nref{2ptlinroll}.\footnote{Here $bQ$ is the same as the exponent $2/(1 - \sqrt{1-1/\Omega} ) $ in equation (6) of \cite{Dubovsky:2008rf}, for $\Omega = Q^2/4$.}  Strictly speaking, the stochastic calculation is an ensemble average, whereas we observe only one realization of the reheating surface. However, for a statistically homogeneous field and a region containing many approximately independent correlation volumes, a spatial average over the reheating surface is expected to provide the corresponding ensemble average over the short-distance modes. For more comments on this point, see section \ref{GMCReh}.  

An observation is that the RG transformation we study in this paper runs in the opposite direction from the usual Starobinsky evolution. Our RG transformation is a Wilsonian coarse-graining of the final reheating surface: starting from the physically regulated reheating geometry, we integrate out short comoving wavelengths and move toward longer comoving scales, i.e.~toward the IR. In the inflationary interpretation, this reconstructs the geometry at earlier and earlier Lorentzian times.

By contrast, in the usual Starobinsky description one evolves forward in Lorentzian time while keeping the coarse-graining scale fixed in physical units. Because the universe is expanding, this fixed physical scale corresponds to progressively shorter comoving wavelengths, so shorter and shorter comoving modes enter the stochastic long-wavelength field. Thus forward Lorentzian evolution corresponds to moving toward the UV of the fiducial theory on the reheating surface. In this sense, the future of the bulk corresponds to the UV of the fiducial theory and the past to the IR, precisely as in the usual UV/IR relation of dS/CFT~\cite{Strominger:2001pn}.

\subsection{Eternal inflation as a breakdown of RG}

In order to determine the precise value of $Q$ where eternal inflation sets in, \cite{Creminelli:2008es} computed the expectation value of the reheating volume produced from a finite initial region. We choose the initial slice to be at  $\tau_{\rm IR}=\tau_0=0$. The local reheating time $\tau(x)$ is then also the duration of the stochastic evolution, and hence the random RG time discussed in section \ref{newrenliouv}.
We are computing 
\begin{equation}
\mE[V]=\int d^{d}x\, \mE[e^{d \,\tau }] = \int d^d x\, e^{d \zeta}\big|_{\rm phys}\,.
\end{equation}
It was observed in \cite{Creminelli:2008es} that this expectation value has the following form for large $\tau$
\be 
\int d\tau  \frac{1}{\tau^{3/2}}\, e^{ - d ( Q^2/4 -1)  \tau} ~,~~~~~~{\rm for } ~~~ \tau \gg 1
\ee 
so that when $Q^2< 4$ the integral diverges.\footnote{Their parameter $\Omega$ is $\Omega = { Q^2 /4}$.} This is interpreted as the onset of slow roll eternal inflation. 
In other words, 
\begin{equation} \la{EIC}
{ \dot \phi^2 \over H^{d+1} } {\rm Vol}(S^{d+1} ) = Q^2 < 4 ~~ \Longleftrightarrow ~~ \text{Eternal inflation}
\end{equation}
This is precisely the  opposite of the  convergence criterion for the renormalization of the volume operator. When the integral converges this renormalization gives the operator $: e^{ d b Q \zeta} :$ with $b$ in \nref{KPZnIntro}. While $b$ becomes complex in the regime of \nref{EIC}, the proper physical interpretation is that the computed expectation value diverges. 

From the point of view of random geometry, this transition marks the breakdown of the ordinary fixed fiducial cutoff description of the volume measure. For $Q>2$, we can describe the problem in terms of a logarithmic conformal field theory on a fiducial manifold. We can say that in this regime, the fiducial geometry emerges in the continuum limit. This occurs in our universe, where we are sensitive to a region that contains many Hubble volumes at the end of inflation.   Indeed, mathematicians have proven that a continuum limit of the random measure exists \cite{Kahane1985} for $Q>2$.  We can similarly say that in eternal inflation, we do not have the emergence of a smooth comoving (or average) geometry at long distances.

There are some discussions of aspects of random geometry for $Q<2$; see \cite{Gwynne:2019bhi}. It would be interesting to determine whether this can help clarify eternal inflation. More generally, it would be interesting to understand if there exists a formulation where one studies the random surface in its own right, without reference to the fiducial geometry. If such a construction existed, it would likely be relevant in the regime of eternal inflation.

As a side comment, note that we have always been discussing the geometry of the reheating surface at various scales. If we want to discuss the long distance geometry of our universe in the present era, then we also need to take into account the detailed cosmological evolution after reheating. One can indeed obtain an effective theory at large scales by integrating out the shorter scales properly \cite{Baumann:2010tm}.

\subsection{ Contributions to correlators for other fields}

We now consider the corrections to a two-point function of a massive field that arise from the fluctuations of the duration of inflation, or the short-distance fluctuations of $\zeta$. 

We consider a field $\chi$ with a mass in the range $ 0 < m/H < d/2 $. In de Sitter space in flat slicing  this gives rise to the late time correlator 
\begin{equation}
\label{2ptbare}
\langle \chi(\tau, x)\chi(\tau', y)\rangle_{\text{QFT}}=\frac{e^{-\Delta_{m} \tau }e^{-\Delta_{m}\tau'}}{|x-y|^{2\Delta_{m}}} \,, ~~~~~~{\rm with }~~~ \Delta_{m}=\frac{d}{2}-\sqrt{\frac{d^{2}}4-\frac{m^{2}}{H^{2}}}\,.
\end{equation}
We should evaluate this correlator on the reheating surface. Since $\phi=\phi_r$ there, equation \nref{ZetPhiRel} implies that the local reheating times are $\tau_{\rm reh}(x) = \zeta(x)$ and $\tau_{\rm reh}(y) = \zeta(y)$, up to the uniform additive constant already absorbed into $\zeta$. Note that \nref{2ptbare} is the two-point function for a given instance of the reheating surface. We now look at the problem at longer distances, with a fixed cutoff in comoving coordinates, and averaging over shorter distances. In other words, we perform an RG transformation and map the problem to a theory with a fixed fiducial cutoff. Then the exponentials of $\tau$ or $\zeta$ change as in \nref{expdeltaa}. Therefore, the two-point function of $\chi$ in the full theory is given by  
\begin{equation} \la{CorrInf}
\langle \chi(x)\chi(y)\rangle=\frac{\langle :e^{-d\alpha_{m}Q\zeta(x)}\!:\,: e^{-d\alpha_{m}Q\zeta(y)} \! : \rangle_{\text{fid}}}{|x-y|^{2\Delta_{m}}}
\end{equation}
with\footnote{In comparing with \nref{KPZhVa} notice that the initial $h$ in \nref{2ptbare} is negative.} 
\begin{equation}
\alpha_{m}=\sqrt{\frac{Q^{2}}{4}+\frac{\Delta_{m}}{d}}-\frac{Q}{2}\,.
\end{equation}

In the context of two dimensional gravity, one often considers integrated operators such as \nref{IntMat}. Note that here, the final operator in \nref{CorrInf} has full conformal dimension zero, with the ``gravitational dressing'' given by $e^{ - d \alpha_{m} Q \zeta}$ canceling the dimension $\Delta_m$ of what we could call a ``matter'' operator. Producing a dimension zero operator is another way to ensure invariance under the conformal coordinates transformations we discussed around \nref{zetaCh}. Of course, 
one must also specify the points $x$ and $y$  in a diffeomorphism-invariant way.
 We imagine doing so
by constructing suitable observables in the later universe that do not introduce further local $\zeta$ dependence at either point. Note that in writing \nref{CorrInf} we assumed that we regulated or averaged the field $\zeta$ over a small region around each point, over a comoving cutoff region $\hat \epsilon \ll |x-y|$.

As a side comment, note that formula \nref{CorrInf} is reminiscent of a similar formula in nearly AdS$_2$ that gives the physical correlator of a boundary operator in terms of the correlator of a bulk field in AdS$_2$ and the boundary reparametrization mode; see equation (4.25) in \cite{Maldacena:2016upp}.

\section{Connections with the mathematical literature}
\label{sec:mathlit}

In this section, we discuss relations between some of the points we brought up in this paper and the mathematical literature on random surfaces and log-correlated random measures. 

\subsection{Quick review of Gaussian Multiplicative Chaos }
\label{qckGMC}

Mathematicians developed a general construction, known as Gaussian multiplicative chaos (GMC) \cite{Kahane1985, Rhodes:2013iua}, in which one starts with a Gaussian field 
$X$ satisfying 
\begin{equation}
\label{Xcov}
\langle X(x)X(y)\rangle=\log\frac{1}{|x-y|}+g(x,y)
\end{equation}
with $g(x,y)$ a function that is regular as $x$ approaches $y$. They then use it to define a random volume measure on a fixed ``fiducial'' manifold via
\begin{equation}
\label{Vgmc}
dV=\,\normord{\,e^{\gamma X}}d\hat{V} ~,~~~~~~~d {\hat V }= d^d x
\end{equation}
with $d\hat{V}$ the volume element of the fiducial manifold \cite{Kahane1985,Rhodes:2013iua}, which we took to be flat space. The construction itself simply defines a random measure, but does not require $X$ to have an interpretation as the Weyl factor of a physical metric.\footnote{One exception is the case of $2d$ Liouville gravity, where the associated GMC measure has this interpretation on firmer footing (see \cite{David:2014eua}).} Our renormalization result in section \ref{renarg} provides an interpretation of this measure as the appropriate volume of a random surface. If we identify our variables with the GMC convention via
\begin{equation}
X=\sqrt{\frac{d}{2}}\,Q\zeta\,,~~~~ \gamma=\sqrt{2d}b\,,~~\text{ or }~Q=\frac{\sqrt{2d}}{\gamma}+\frac{\gamma}{\sqrt{2d}} \la{VolMath}
\end{equation}
we recover
\begin{equation}
\la{VolUsGMC}
dV=\,\normord{\,e^{\gamma X}}d\hat{V}=\,\normord{\,e^{dbQ\zeta}}d\hat{V}
\end{equation}
which for $Q > 2$ is precisely the volume element of the metric \nref{metric} written at a fixed fiducial cutoff. Our interpretation would also hold in a generalization of GMC where we only assume that at short distances the field is effectively Gaussian and log-correlated. A warning is that the random surface theory that implies a given covariance \nref{Xcov} might generically be described by a non-local action. In the inflationary application from section \ref{sec:inflation}, this non-locality is not a surprise, and it arises naturally through de Sitter-like evolution in an extra dimension.

We should note that there is also literature on ``supercritical GMC'', which in our conventions corresponds to the real branch $b>1$. In this regime, modified GMC constructions can produce non-trivial ``atomic'' measures \cite{Barral:2012utw}. This is a distinct regime from the one studied in this paper, since the $b > 1$ solution at $Q > 2$ corresponds to the second branch of the volume operator and does not have the correct semiclassical limit. By contrast, for $Q < 2$, we have that $b$ becomes complex rather than continuing onto a real branch with $b > 1$.

A relevant point about GMC, which we will use later, is that the volume measure \nref{Vgmc} gives full measure to so-called ``thick points'' \cite{Kahane1985, Rhodes:2013iua}. Thick points are points $x$ at which the regulated field $\zeta_{\hat \epsilon}(x)$ grows linearly with $\log \hat \epsilon^{-1}$ as the fiducial cutoff $\hat \epsilon$ is taken to zero. To be more specific, an $a$-thick point is defined, in our notation, as a point $x$ where
\begin{equation}
\text{lim}_{\hat \epsilon \rightarrow 0} \frac{\zeta_{\hat \epsilon}(x)}{\log \hat \epsilon^{-1}}=a \,.
\end{equation}

From the perspective of the volume $\hat{V}$ of the fiducial background $d\hat{s}^{2}$, the set of thick points has measure zero. To see this, remember that for a finite, small cutoff  we can treat $\zeta_{\hat \epsilon}(x)$  locally as a Gaussian with variance
\begin{equation}
\label{zetavar}
\langle \zeta_{\hat \epsilon}^{2}(x)\rangle\approx \frac{2\tau}{dQ^{2}} ~,~~~~~~~~{\rm with}~~~\hat \epsilon =e^{-\tau}\,.
\end{equation}
This implies that an $a$-thick point is exponentially suppressed in $\tau$ as
\begin{equation}
\label{pthick}
\zeta_{\hat \epsilon}\sim a \tau
~~\longrightarrow~~
p\sim e^{-\frac{\zeta_{\hat \epsilon}^{2}}
{2\langle\zeta_{\hat \epsilon}^{2}\rangle}}
= e^{-\frac{d a^{2}Q^{2}}{4}\tau}\,.
\end{equation}
Therefore, in the continuum limit $\tau \rightarrow \infty$, a point sampled uniformly from the fiducial manifold has zero probability of being a thick point. In fact, we can estimate the Hausdorff dimension of the set of $a$-thick points in an intuitive way, by imagining that we decompose the manifold into balls of fiducial radius $\hat{\epsilon}$. There are then $\hat{\epsilon}^{-d}$ such balls in the manifold, each of which has probability \nref{pthick} of being an $a$-thick point. It follows that the expected number of $a$-thick points is
\begin{equation}
N_{a} \sim \hat{\epsilon}^{-d}\,\hat{\epsilon}^{\frac{da^{2}Q^{2}}{4}}=\hat{\epsilon}^{-d\big(1-\frac{a^{2}Q^{2}}{4}\big)}
\end{equation}
which implies that the set of $a$-thick points has Hausdorff dimension
\begin{equation}
\label{dathick}
d_{\text{a-thick}}=d\bigg(1-\frac{a^{2}Q^{2}}{4}\bigg)
\end{equation}

While our argument was heuristic, mathematicians have shown \nref{dathick} rigorously \cite{HuMillerPeres2010, Cipriani:2017aqo}. Another interesting thing they proved is that for $a \leq \frac{2}{Q}$ the set of $a$-thick points is dense, e.g,  any open set in the fiducial manifold contains such thick points.\footnote{More specifically, these results were established in $d=2$ and $a \leq \frac{2}{Q}$ in \cite{HuMillerPeres2010}. For $d>2$, the results were only established for $a<\frac{2}{Q}$, in \cite{Cipriani:2017aqo}}

However, if we sample points using the actual measure $dV \propto e^{dbQ\zeta}d\hat{V}$ of the random surface, thick points are, in fact, generic. The reason is that while they are unlikely, they give an exponentially large physical volume contribution, due to the exponential of $\zeta$ in \nref{Vgmc}. An intuitive explanation for this fact comes from combining the rarity of an $a$-thick fluctuation with the volume weight assigned to it.  At finite fiducial cutoff, this combined probability weight is
\begin{equation} \la{ThickEst}
P(\zeta) :e^{dbQ\zeta}:\big|_{a\text{-thick}}\sim e^{dbQa\tau} e^{- d b^2 \tau}e^{-\frac{da^{2}Q^{2}}{4}\tau}=e^{-\frac{dQ^{2}\tau}{4}\big(a-\frac{2b}{Q}\big)^{2}}
\end{equation}
where we have taken to the fiducial cutoff to be $\hat \epsilon=e^{-\tau}$ as before, and $P$ denotes the Gaussian probability weight estimated in \nref{pthick}. The second factor on the left-hand side contains the normal-ordering subtraction of the operator.  We see that this combined quantity \nref{ThickEst} is maximized for thick points with 
\be 
\la{VolThi} 
a=\frac{2b}{Q} ~.
\ee 
Moreover, the distribution becomes sharply peaked as $\tau \rightarrow \infty$.   
Indeed, mathematicians have shown \cite{Kahane1985, Rhodes:2013iua} that in the continuum limit, the set of thick points with $a=\frac{2b}{Q}$ carries all the physical volume of the random surface.\footnote{In other words, the complement of this set has zero renormalized volume.}.

An interesting statement is then that, as we discussed in \nref{dathick}, from the perspective of the fiducial background $d\hat{s}^{2}$, the renormalized volume is carried by a set of fiducial measure zero, with Hausdorff dimension given by
\begin{equation}
\label{dthick}
d_{\text{fid}}=d(1-b^{2})\,.
\end{equation}
We can understand \nref{dthick} as a statement of how the physical volume of the random surface is distributed when viewed in fiducial coordinates. At finite fiducial cutoff, the Hausdorff dimension statement \nref{dthick} should be weakened to mean that the volume would appear to be clumped around high-field regions across the coordinate system. Furthermore, as we vary $b$ from $0$ to $1$, the volume would be spread less and less uniformly, until it becomes effectively carried by a point-like set of zero Hausdorff dimension at $b=1$. Note that since $b=1$ corresponds to $Q=2$, this threshold corresponds to just before the regime where the smooth description of the random surface breaks down.\footnote{The case $b=1$ is subtle. At this critical value, the usual GMC normalization gives a trivial continuum limit, while a non-trivial ``critical measure'' can be obtained with a modified renormalization \cite{Duplantier:2012saa}.}

We now turn to connect these notions to inflation.

\subsection{GMC as a theory of the reheating surface}
\la{GMCReh}

The reheating surface in linear roll inflation is described by a random surface theory in the general class considered in this paper. Therefore, from the general point of section \ref{qckGMC}, the volume of the reheating surface in linear roll inflation is described by a GMC measure! This is an interesting connection, because GMC is a very well-studied subject \cite{Rhodes:2013iua}, and one can wonder if it has anything to teach us about inflation. 

In this section, we present a few interesting points in this direction. While they will not be immediately useful for observations, they will give us insights into the random surface behavior of the reheating surface and how random surface properties are related to the dynamics of inflation. 

For example, in section \ref{qckGMC} we explained that, in the continuum limit, the physical volume of the random surface in GMC gives full measure to the thick points of thickness \nref{VolThi}. This means that, in an idealization with a parametrically long period of inflation, a point sampled from the reheating volume measure is associated with a thick point fluctuation of this characteristic thickness.

Thick points are intimately related to fluctuations of the inflaton that slow down the downhill motion of the inflaton. We can see this as follows. With a fiducial cutoff $\hat \epsilon = e^{-\tau} $, a thick point of thickness $a $ behaves as $\zeta \sim a \tau $. Using \nref{ZetPhiRel}, and recalling the connection between Lorentzian time and the RG time $\tau$,  this then means that 
\be \la{PhiEv}
{H \over \dot \phi } \left[ \phi(x,\tau) - \phi_r \right]  = (1-a) \tau   + {\rm constant}\,.
\ee 
Note that the usual classical downhill evolution would have $a=0 $ in the right-hand side of \nref{PhiEv}.  We have said that the volume is dominated by thick points of thickness \nref{VolThi}, meaning that a volume-typical observer will think that the inflaton had a downhill velocity given by 
\be \la{PhiEvt}
{H \over \dot \phi } \left[ \phi(x,\tau) - \phi_r \right]  = \left(1- { 2 b \over Q } \right) \tau  + {\rm constant}\,.
\ee  
Here the microscopic drift $\dot \phi$ in the stochastic equation is not changed; rather, the accumulated stochastic fluctuations partially cancel the classical drift along these rare conditioned trajectories

We now rederive these points in a slightly different language. 
Let us re-examine the volume renormalization we discussed in section \ref{renarg} from a different perspective. Let us say that we are studying the reheating surface in inflation with a fixed physical cutoff, but we only know the $\zeta_{\text{IR}}$ modes of the regulated $\zeta_{L}=\zeta_{\text{IR}}+\zeta_{\tau}$, with notation as in section \ref{renarg}. Remember that $\zeta_{L}$ is the amount of inflation since the modes in $\zeta_{\text{IR}}$ crossed the horizon. We assume $\zeta_{L}$ will be typically large. Conditioning on $\zeta_{\text{IR}}$, we can ask what $\zeta_{\tau}$ is at volume-typical points of the reheating surface. 

From the discussion of GMC, we know that these volume-typical points should be thick points with a given thickness. For a fixed physical cutoff, the local fiducial cutoff scales as $e^{-\zeta_L}$ up to a fixed multiplicative constant. Thus, in the limit of large RG time, the thick-point relation with thickness $2b/Q$ implies
\begin{equation}
\label{thickzeta}
\zeta_{\tau} \sim \frac{2b}{Q}\,\zeta_{L}
\,\,\longrightarrow\,\, \zeta_{\tau} \sim \frac{2b}{Q-2b}\,\zeta_{\text{IR}}\,, \qquad \zeta_{L} \sim \frac{Q}{Q-2b}\,\zeta_{\text{IR}}\,.
\end{equation}

Note that this is a result we also obtained from explicit averaging over paths in section \ref{simpcom}. 
Here, we are showing instead that these leading configurations $\zeta_{\tau}$ are precisely the configurations that are typical when points are sampled according to the physical volume measure. This explains more directly why they dominate a typical realization of the random surface: spatial averaging over a sufficiently large realization produces the same effect as the coarse-grained ensemble average. Moreover, note from \nref{thickzeta} that the short-distance $\zeta_{\tau}$ and $\zeta_{\text{IR}}$ are linearly related. 

From inflation, this means that from the time that $\zeta_{\text{IR}}$ crossed the horizon, the inflaton was rolling down with an average drift effectively smaller than the classical one. Classically, the rolling time would be $\zeta_{\text{IR}}$ itself, but the actual rolling time $\zeta_{L}$ is bigger by a factor of $\frac{Q}{Q-2b}$, as we have shown in \nref{thickzeta}. In other words, this means that the inflaton trajectory leading to a thick point is effectively slower by a deterministic amount, given by
\begin{equation}
\label{thickphi}
\dot{\phi}\big|_{\text{thick point}}=\bigg(1-\frac{2b}{Q}\bigg)\dot{\phi}
\end{equation}
which is equivalent to \nref{PhiEvt}.

In particular, the volume-typical points in the reheating surface correspond to these slower inflaton trajectories. As we take the number of e-folds to be very large, the probability of these slower trajectories is exponentially suppressed, but they nevertheless dominate the computation, as expected from thick points. From \nref{thickphi}, we see that the configurations near the semiclassical limit (large $Q$) are only slower by a small amount, but as $Q \rightarrow 2$, the dominating configurations become those where the inflaton was barely moving, and whose volume in the future is very large. Along these trajectories, the upward quantum fluctuations of the inflaton are canceling its classical motion.

We should also comment that right before the eternal inflation transition, at $Q=2$, $d_{\text{fid}}$ \nref{dthick} discussed in section \ref{qckGMC} goes to zero, which indicates that the volume-carrying set becomes point-like in the fiducial coordinates. However, the role of thick points at a finite cutoff, and thus the validity of our analysis for inflation, becomes unclear there since the inflaton field at these thick points would have zero drift from \nref{thickphi}. This signals that the volume measure is increasingly dominated by exceptionally long inflationary trajectories. At $Q<2$ we cannot use the volume element with a fixed fiducial cutoff, so the current analysis would not make sense.

As we mentioned, ``thick points'' are rare fluctuations where the scale factor of the metric becomes unusually large. Rare fluctuations of a similar kind have been discussed as a way of generating primordial black holes, for a suitable range of parameters. For a review, see \cite{Byrnes:2021jka}.

\subsection{Uphill inflaton fluctuations as thick points}

In section \ref{renarg} we argued that the expectation of the volume $V$ of the random surface, at a fixed fiducial cutoff, diverges for $Q<2$. A natural question is whether the expectation value of different powers $V^{n}$ diverges at the same value of $Q$.  One can answer this straightforwardly by looking for UV divergences of $V^{n}$, see \nref{Multiple}, 
\begin{equation}
\label{ncorrel}
\langle V^{n}\rangle=\int \prod_{i=1}^{n}d^d x_i\,\langle \normord{\,e^{dbQ \zeta(x_{1})}} \cdots \normord{\,e^{dbQ \zeta(x_{n})}} \rangle
\end{equation}
explicitly in the free-field approximation. This is reasonable if, at short distances, $\zeta$ is Gaussian and log-correlated, which implies that the contribution of $n$ points at a distance of at most $r$ from each other is
\begin{equation}
\langle V^n \rangle \supset\int \prod_i d^dx_i  \prod_{i> j} |x_i-x_j|^{ - 2 d b^2 } \sim  \int {dr \over r} \,r^{d (n-1) (1-b^2 n)}\,.
\end{equation}
This means that the $n$-th power of the volume diverges for  
\begin{equation} \la{nbsq}
nb^{2} \geq 1\,.
\end{equation}

Remember that as we take $Q$ from infinity to $Q=2$, $ b$ goes from $0$ to $b=1$ \nref{KPZnIntro}. Therefore, larger $n$ diverges for larger values of $Q$. Also, for any given $Q$, there is always a sufficiently large $n$ for which $\langle V^{n}\rangle$ diverges. This was noted before for the inflation case in \cite{Creminelli:2008es} and in this form for Liouville in \cite{Chatterjee:2024phq}. 

Another way of thinking about the computation that makes the discussion somewhat reminiscent of thick points is that we can consider the contribution to \nref{ncorrel} from the small $r$ region as coming from the leading OPE of the $n$ operators. Keeping only the wavelengths bigger than $r$ in these operators, which are their shared modes, their product can be approximated as
\begin{equation}
\label{nope}
\normord{\,e^{dbQ \zeta(x_{1})}} \cdots \normord{\,e^{dbQ \zeta(x_{n})}} \,\approx e^{ndb Q\zeta_{r}}r^{ndb^{2}}
\end{equation}
where $\zeta_{r}$ only has wavelengths bigger than the fiducial size $r$, and the last term is the subtraction associated with this cutoff. Because of the exponential in \nref{nope}, the same argument as in section \ref{qckGMC} implies that at small $r$ the configurations maximizing \nref{nope} will scale as log $r^{-1}$.  We can maximize the contributions to \nref{nope} of configurations with $\zeta_{r} \sim a \log r^{-1}$  
\begin{equation}
\la{ThickCon}
\normord{\,e^{dbQ \zeta(x_{1})}} \cdots \normord{\,e^{dbQ \zeta(x_{n})}}|_{\text{a-thick}}\sim   r^{\frac1{4}d a^2 Q^2} \times  r^{-d n b Q a} \times r^{ d b^2 n}=r^{\frac{dQ^{2}}{4}\big(a-\frac{2nb}{Q}\big)^{2}}r^{-db^{2}n(n-1)}
\end{equation}
where the first factor is the probability \nref{pthick} that we have such a thick point, and the last two follow from \nref{nope}. We see therefore that the contribution is maximized for thick-point-like configurations that scale with $r$ as 
\begin{equation}
\label{zetan}
\zeta_{r}=a \log r^{-1}    ~,~~~~~~~a = { 2n b \over Q}\,.
\end{equation}
Multiplying \nref{ThickCon} by the extra fiducial volume suppression factor $r^{ d(n-1)}$ from $n$ points coming close together, we recover the divergence when \nref{nbsq} is obeyed. We note that in the regime $nb^2 > 1$  where the moments diverge, the optimal value of $a$ satisfies $a = \frac{2nb}{Q} > \frac{2}{Q}$, and thus these configurations should not be interpreted as ordinary thick points that persist in a typical continuum realization. We can think of them as finite-scale large deviations with thick-point-like scaling, which dominate the divergence of high moments.

In the context of inflation, it is interesting to understand what these configurations look like from the point of view of the inflaton evolution. Indeed, \cite{Creminelli:2008es} perform essentially the same computation of $\langle V^{n}\rangle$ from their stochastic inflation model. In that context, the contributions we discussed come from very long inflationary evolutions, where initially the $n$ points were very close together. 

\begin{figure}[t!]
    \centering
    \includegraphics[width=\linewidth]{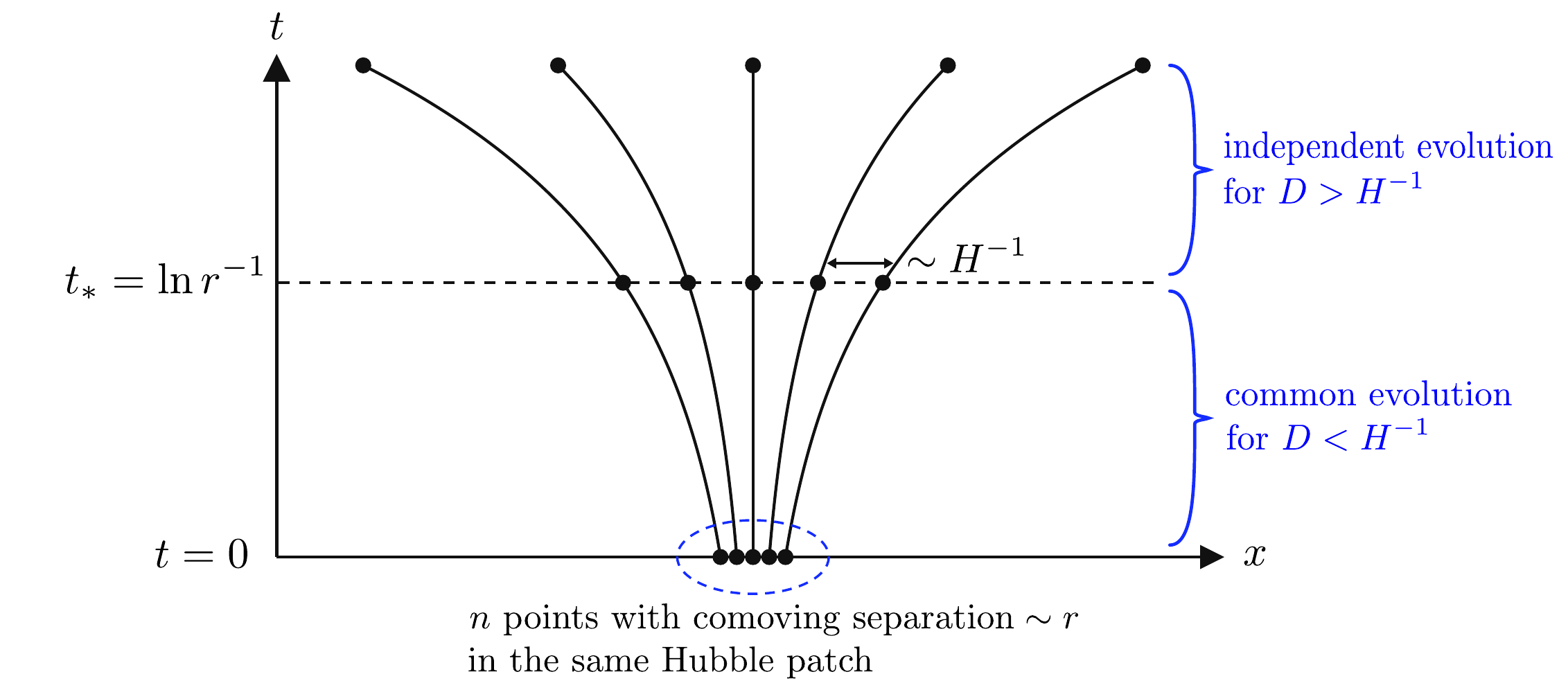}
    \caption{Dynamics of five points that are initially close together on the initial-condition surface. We use $D$ to denote a typical distance between the points at a given time. For $t < t_{*}$, the points lie within the same Hubble patch and therefore share the same inflaton dynamics. For $t > t_{*}$, they are separated by more than a Hubble radius, and their subsequent inflaton dynamics can be treated as independent.}
    \label{splitfig}
\end{figure}

They approximate the evolution into two parts: In the first part, the $n$ points were within the same Hubble patch, such that the regulated inflaton is the same at all points. Then, at a time $\tau_{*}=\ln r^{-1}$, the $n$ points separate and become further apart than Hubble, after which we have the second part where we treat their evolution as completely independent (see Figure \ref{splitfig}), until the end of inflation.  This is just a spacetime realization of the OPE argument we have discussed.  

The period of common evolution corresponds to the modes of wavelength larger than $r$ shared by the field at the $n$ different points, which is manifested as the fact that the inflaton evolution at these points is the same. The subsequent evolution corresponds to shorter-wavelength modes, which are effectively independent at the $n$ points, and just renormalize the operators, changing $e^{d \zeta} \to:e^{ d b Q \zeta}:$.

It is interesting to note that in the paper they discuss that what makes $V^{n}$ diverge are ``uphill inflaton fluctuations'', which are very large deviations from classical motion that scale with $\tau_{*}$ in the common evolution. They are very unlikely, but make the volume in the future much larger. Indeed, these ``uphill inflaton fluctuations'' are precisely the thick point-like configurations discussed in \nref{zetan}!

We can see that they are ``uphill fluctuations''   during the common evolution as follows. Note that for the thick point-like configurations we discussed, $\zeta_{r} =a \log r^{-1}=a \tau_{*}$, with $\tau_{*}$ the time the points separated in the bulk, see Figure \ref{splitfig}, and $a $ in \nref{zetan}. By also using \nref{nbsq} we conclude that $a>1$.  Now, looking at \nref{PhiEv}, during this period of time, we conclude that the inflaton is moving in the opposite direction it would have moved classically (for $a=0$), therefore we are having an ``uphill'' fluctuation. 

So far, we have been discussing the strict linear roll model. In \cite{Creminelli:2008es} they modify the inflaton potential in a way which would suppress large uphill inflaton fluctuations. As we just discussed, this is similar to saying that we only allow thick points of thickness $a\leq 1$. With this restriction, one can then repeat the calculation around \nref{ThickCon} and find that all powers $\langle V^n \rangle $ diverge only at $Q=2$, as discussed in \cite{Creminelli:2008es}.

It was also argued in \cite{Creminelli:2008es} that, when $Q<2$, there is a positive probability for the physical volume to be infinite. More precisely, a finite fiducial region with finite coarse-grained $\zeta_{\rm IR}$ can nevertheless have a positive probability of evolving to a reheating surface with infinite physical volume. 

\subsection{Discrete tilings and discrete RG}
\label{dyadsqr}

In this section, we discuss connections between our renormalization argument in section \ref{renarg} and related work in the mathematical literature on random surfaces.  

\begin{figure}[t!]
    \centering
    \includegraphics[width=0.8\linewidth]{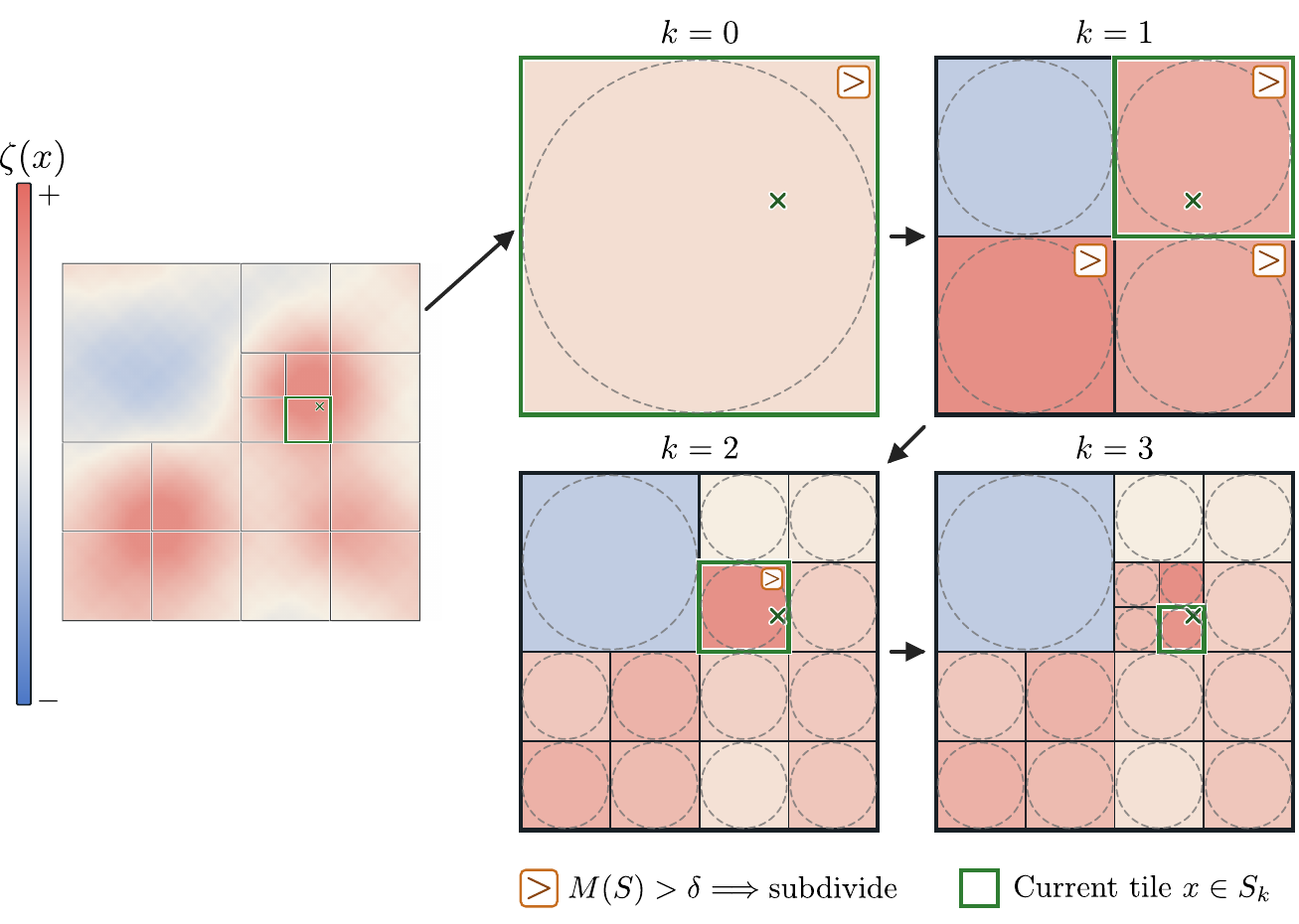}
    \caption{On the left, we have the plot of the continuum field across an entire square, with field value indicated by color. Each delimited square inside it is a fundamental tile, according to the definition of \cite{Gwynne:2019bhi}. On the right, we show the process of splitting the original square to obtain these tiles. At each step of the split, the squares are treated as having a constant field value, indicated by their color. This is the value of the field averaged over the dashed circle inside the square. One starts with the entire square at $k=0$. If $M>\delta$ for a square, it is split into $4$ daughter squares. This continues until all squares satisfy $M<\delta$. Note that as the squares get smaller, the circle average better resolves the continuum field. From the perspective of a marked point $x$, the LQG size of its current square $M$ is undergoing a random walk in discrete time $k$, which stops once $M$ becomes less than $\delta$ for the first time. 
    \label{splitsquare}}
\end{figure}

The first work we discuss is \cite{Duplantier:2008prc}, which, in the context of $2d$ Liouville gravity, was interested in relating the usual fractal dimension of sets in a fiducial manifold to a ``quantum'' notion of fractal dimension they define (for a quick review, see section \ref{fracdim}). However, the main connection\footnote{Another less direct connection is that they find a first passage problem similar to the one we discussed in section \ref{renarg}. However, the calculation and settings are different; in particular, they are already using the volume element at a fixed fiducial cutoff in their calculation, which is what we wanted to derive in the first place on section \ref{renarg}.} of their paper to ours is that they introduce a tiling of the manifold into dyadic squares $S$ (dyadic just means that the squares have side lengths which scales like $2^{-n}$, see Figure \ref{splitsquare}) of approximately fixed Liouville volume, defined by \nref{VolUsGMC}. Since the Liouville volume corresponds to an actual physical volume, from our perspective this construction has the potential to provide a discretization of the geometry into fixed physical size tiles. One issue is that the discretization would use the exact continuum Liouville volume, which is a very fine-grained quantity that is not even well defined for $Q<2$.  

However, in the paper \cite{Duplantier:2008prc}, they also introduce a more coarse-grained quantity that approximates well the dependence of the volume of a square on its fiducial side length $r$, which in our notation is written as
\begin{equation}
\label{Vsheffdup}
V(S)\sim (re^{\zeta_{r}(v_{S})})^{2bQ}= r^2 : e^{ d b Q \zeta_r } :
\end{equation}
where $\zeta_{r}$ is supposed to be the field $\zeta$ but only with wavelengths larger than the fiducial scale $r$, evaluated at the center $v_{S}$ of the square. Note that equation \nref{Vsheffdup} is the renormalized volume at fixed fiducial cutoff, with the cutoff set to be the scale $r$ of the square. 
The approximation \nref{Vsheffdup} also makes manifest that the volume of the square is fixed by a physical length scale $r e^{\zeta_{r}}$.

A discretization of the manifold using the more coarse-grained quantity \nref{Vsheffdup} was later pursued in \cite{Gwynne:2019bhi}, which is perhaps the paper most related to our work. To be more specific,  in the paper, the authors propose the same dyadic square tiling of the manifold, but instead of using the volume to divide tiles, they used a quantity $M(S)$, which they called LQG-size, defined in our notation as 
\begin{equation}
\label{Mdef}
M(S)=(r e^{\zeta_{r}(v_{S})})^{Q}
\end{equation}
with $\zeta_{r}$ evaluated at the center $v_{S}$ of $S$.\footnote{More specifically, $\zeta_{r}$ is defined by averaging $\zeta$ around a circle of radius $r/2$ centered in the middle of the square; see Figure \ref{splitsquare}} Their motivation for doing so was that $M(S)$ is an object that transforms covariantly under coordinate transformations, and whose power $M^{2b}$ approximates the Liouville volume
via \nref{Vsheffdup}. Also, $M$ seems well-defined even at $Q<2$. From our perspective, $M$ is a simple power of the physical scale $r e^{\zeta_{r}}$ of the problem. A non-trivial aspect of the definition \nref{Mdef} is the following. As one makes the squares smaller, there are more modes affecting $M(S)$ since the cutoff of $\zeta_{r}$ decreases. These could make $M$ increase as we decrease $r$, due to large fluctuations. This would not be possible for a physical volume of the manifold, so $M^{2b}$ cannot be interpreted as an exact microscopic volume.

In more detail, their discretization of the manifold works as follows (see Figure \ref{splitsquare}): First, they start with a dyadic tiling of the manifold. For the unfamiliar reader, a dyadic tiling consists of a tiling by squares of side length $2^{-k}$, where $k$ is an arbitrary integer. So, each square has daughter squares with size smaller by a factor of $2$, and a parent square bigger by the same factor. Then, they define a fundamental tile $S$ by the requirement that $M(S)<\delta$, and that all the ancestors $S'$ of $S$ satisfy $M(S')>\delta$. The problem of finding the fundamental tile $S$, containing, for example, a given point $x$, is a first-passage problem, defined as follows: Each square has fiducial side length $2^{-k}$. We can think of a unit increase in $k$, which is going one generation down, as a discrete RG-time step. 

As we increase this time variable, the fiducial length of the square decreases in a deterministic way, and new modes enter $\zeta_{r}$. They treat the way that each new mode enters essentially in a similar way to us in section \ref{renarg}, since they use that at small scales $\zeta_{r}$ behaves as a Gaussian with variance set by $r$. Therefore, the condition that $S$ is a fundamental tile is a discretized version of our first-passage problem in section \ref{renarg}. Furthermore, we can think of the fixed LQG-size tiles of \cite{Gwynne:2019bhi} as being a discretized analog of our idea of regularizing the theory with a fixed physical cutoff. Indeed, an important consistency point, respected by their discretization, is that they allow the field $\zeta_{r}$ to have more modes as they shrink the squares. Therefore, $\zeta_{r}$ at a fundamental tile consistently has modes all the way to the ``physical cutoff'', which is related to the fixed LQG size. 

Moreover, they mention that when $Q<2$, with probability one there is a dense set of ``singularities'', which are points where the fixed LQG size condition $M(S)<\delta$ never truncates. These are related to the fact that the log-correlated field $\zeta$ has a dense set of thick points, as we reviewed in section \ref{qckGMC}. For $Q<2$ there are surely thick points with, in the notation of section \ref{qckGMC}, thickness $a>1$. Around such points, the LQG size $M$ would never become smaller than $\delta$, since $M$ keeps increasing as we decrease $r$. In section \ref{renarg}, we argued that the renormalization relating the theory at a fixed physical cutoff to a fixed fiducial one breaks down at $Q<2$, since the renormalization step fails for the volume element. We can interpret the result of \cite{Gwynne:2019bhi} then as a discrete version of this breakdown, where one cannot realize cells of fixed physical cutoff using finite cells in the fiducial manifold. In the context of inflation, this dense set of ``singularities'' is just the set of points where inflation never ends. See \cite{Gwynne:2019bhi} for other interesting mathematical results about the random surface for $Q<2$.

\subsection{Fractal dimension}
\label{fracdim}

There are different notions of fractal dimension. Here we just review a few that have been discussed in the literature of random surfaces, especially emphasizing the ones that are relevant in our context.  

\paragraph{KPZ quantum dimension.}

 As originally formulated in \cite{Knizhnik:1988ak}, the usual scaling dimension of primary CFT operators is related to a new ``quantum'' scaling dimension after they are coupled to $2d$ gravity. This notion of quantum scaling dimension is mathematically the same as one proposed later in \cite{Duplantier:2008prc} (see also \cite{David:2008su}), which is sometimes called the ``geometric KPZ relation''. 
This is defined by relating the box-counting dimension of fractal sets defined using the fiducial volume versus the physical volume. More precisely, we start from a fractal set $X$ whose box-counting dimension in the fiducial coordinates is $d\,h$. Equivalently, the number of fiducial boxes of volume $\hat\epsilon^d$ needed to cover $X$ scales as
\begin{equation}
N_{\hat \epsilon}(X) \sim (\hat \epsilon^{d})^{-h}\,.
\end{equation}
Then, \cite{Duplantier:2008prc} introduces a quantum box counting dimension defined similarly, but using boxes of fixed Liouville, or ``quantum'', volume $\delta$. Namely, volumes computed using the measure $dV =:e^{ d b Q \zeta}: d^d x $. They show that the number of quantum boxes $\mathcal{N}_{\delta}(X)$ needed to cover $X$ scales as
\begin{equation}
\mathcal{N}_{\delta}(X)\sim \delta^{-\frac{\alpha_{h}}{b}}  
\end{equation}
where $\alpha_{h}$ and $b$ are defined as in our main text. However, if we take $X$ to be the entire random surface, then $h=1$ and both box-counting dimensions are trivial. Therefore, while their result is very interesting, we cannot use it to assign a fractal dimension to the random surface.

\paragraph{$\zeta$-scaling fractal dimension.}

In our introduction, we discussed a different notion of fractal dimension that was based on coarse graining the field $\zeta$. Say we have a torus in the $x$ coordinates,  $x^i = x^i +1$. We can now average $\zeta$ over the torus and thus define the zero mode $\zeta_0$. We can then ask how the volume of the space depends on $\zeta_0$. One would naively think that this dependence should be $e^{ d\zeta_0}$, which is what we expect for a dimension $d$ space. However, we know from the discussions in section \ref{newrenliouv} and \nref{stocrg} that the volume actually scales as $e^{d b Q \zeta_0} $ due to the effects associated with the UV cutoff.

This scaling has a direct analog in the inflationary discussion of \cite{Dubovsky:2008rf}.  There, one specifies a coarse-grained initial configuration and asks how the volume of the eventual reheating surface depends on that initial condition.  In our language, $\zeta_0$ plays the analogous role of a long-wavelength background mode: shifting $\zeta_0$ changes the overall physical size of the geometry, while the shorter-wavelength fluctuations are averaged over.  The non-trivial result is that, after this averaging, the final volume scales as $e^{d b Q \zeta_0}$ rather than the na\"{i}ve $e^{d \zeta_0}$.\footnote{To compare to the notation in \cite{Dubovsky:2008rf}, we would look at equation (6) of \cite{Dubovsky:2008rf} and identify $N_c $ with $\zeta_0$ and $\bar N $ with $bQ \zeta_0$ (and $\Omega = Q^2/4$).}  We can therefore use the response of the volume to a uniform shift of $\zeta_0$ to define an effective $\zeta$-scaling dimension,
\be \la{zetDimF}
\tilde d = d b Q \,.
\ee 
This is the definition that we used to compute \nref{DimUni} for the real world at long distances. 

\paragraph{Volume growth dimension and related ones.}

There is a simple notion of fractal dimension for random surfaces, which is how the volume $V$ of a ball $B_{r}$, of proper (geodesic) radius $r$, scales with $r$. This is the volume-growth dimension $d_{g}$ defined by
\begin{equation}
V(B_{r})\sim r^{d_{g}}
\end{equation}
for small $r$. This perhaps is the simplest and most intuitive definition of fractal dimension. However, determining $d_{g}$ is non-trivial, since the points up to distance $r$ are defined from a geodesics problem on the actual random surface, with metric $ds^{2}=e^{2\zeta}d\hat{s}^{2}$. For example, since $\zeta$ is random, the geodesics can find ``shortcuts'' by exploring ``valley'' regions where $\zeta$ has a very negative fluctuation. Indeed, even in the most understood random surface theory, which is $2d$ Liouville gravity, $d_{g}$ is only known for $Q^{2}=\frac{25}{6}$, which is the pure gravity case. In that case, it is known that $d_{g}=4$, thanks to the connection to random planar maps  (or planar diagrams) \cite{Ambjorn:1995dg,LeGall2007,LeGall2013}.  

The study of $d_{g}$ in $2d$ Liouville gravity is an active field, with interesting new developments. There is a candidate distance function, which naturally generalizes to higher $d$, defined as follows \cite{Ding:2018uez}
\begin{equation}
\label{Dgeo}
D(x,y)=a(\hat\epsilon)\text{ inf}_{\gamma} \bigg(\int_{\gamma} d\hat{l}\, :e^{\xi Q \zeta_{\hat \epsilon}} : \bigg)
\end{equation}
where $d\hat{l}$ is the length element of the path in the fiducial metric, and the $\text{inf}$ symbol stands for minimization over paths from $x$ to $y$. Moreover, $\zeta_{\hat{\epsilon}}$ has fixed fiducial cutoff $\hat\epsilon$, and $a$ is an appropriate subtraction term. A non-trivial aspect is the $Q$-dependent exponent $\xi$, which is necessary for self-consistency and is hard to compute \cite{Ding:2018uez, Ding:2019ecf, Gwynne:2019bhi, gwynne2020existenceuniquenessliouvillequantum}, even for $d = 2$.

 Once we know $\xi$, we can easily find $d_g$ as follows. By rescaling $\zeta \rightarrow \zeta+\lambda$, the distances are rescaled by $e^{\xi Q \lambda}$, and the volumes by $e^{dbQ\lambda}$, so by consistency
\begin{equation}
d_{g}=\frac{d\,b}{\xi}\,.
\end{equation}
Therefore, if one knew $\xi$ perturbatively at large $Q$, we could compute the volume growth dimension $d_{g}$ of the reheating surface in our universe! This is not a sharp observable quantity, but it is an interesting calculation. In analogy with the KPZ formula \nref{kpzapprox}, one would have expected the first correction to $d_{g}$ to go as $Q^{-2}$. However, in $d>2$, there are known upper and lower bounds \cite{ContrerasHipZhuang2025bounds, DingGoswami2019upper, ContrerasHipZhuang2025weak} on $d_{g}$ at large $Q$, which imply that this cannot be the case. In particular, it is known that at large $Q$\footnote{Note that mathematicians have a slightly different definition of $Q$ at $d>2$ than we do (see for example \cite{ContrerasHipZhuang2025bounds}). They are related via $Q\big|_{\text{math}}=\sqrt{\frac{d}{2}}\,Q\big|_{\text{ours}}$, as one can see by using \nref{VolUsGMC}.} (or small $b$)
\begin{equation}
2\sqrt{d(d-1)}Q^{-1}>d_{g}-d>\frac{c_{d}Q^{-\frac{4}{3}}}{\log Q}~~,~~ Q \gg 0
\end{equation}
with $c_{d}$ a positive constant. This suggests that the first-order correction to $d_{g}$ must come not from simple perturbation theory in $Q^{-2}$, but instead from a more complicated resummation.

We should also mention two other related notions of fractal dimension for a random surface. We do not treat them as fully different, because one would intuitively expect them to match $d_{g}$ for random surfaces that are well behaved enough. One of these notions is the Minkowski, or box counting dimension, $d_{M}$. It is defined from the minimum number $N(r)$ of geodesic balls of radius $r$ needed to cover the manifold, that is
\begin{equation}
d_{M}=\lim_{r \to 0}\bigg[\frac{\log N(r)}{\log r^{-1}}\bigg].
\end{equation}

Intuitively, if the manifold has volume $V$, one would expect $V/r^{d_{g}}$ balls of radius $r$ to be necessary to cover it. So, intuitively $d_{M}$ should be $d_{g}$. A second notion of fractal dimension is the Hausdorff dimension $d_{H}$, which, for similar intuitive reasons, one would expect to match $d_{M}$, and therefore $d_{g}$. We should mention that in the context of the distance function defined by \nref{Dgeo} in $d=2$, the three fractal dimensions $d_{g}$, $d_{M}$ and $d_{H}$ were shown to be equal\cite{Gwynne:2019ome, Ang:2020kke}.

We now would like to make a conjecture involving the distance formula \nref{Dgeo}, in analogy to our discussion for the formula for the volume in section  \nref{renarg}. We conjecture that \nref{Dgeo} arises from the distance defined using the physical metric and a physical cutoff $\epsilon$, after we integrate out shorter distances. In other words, 
\begin{equation}
\label{Dprop}
D(x,y)=G(L)\bigg\langle \text{inf}_{\gamma_p} \bigg(\int_{\gamma_p} d\hat{l}\,e^{\zeta_{L}}\bigg)\bigg\rangle_{\!\!\text{short}}
\end{equation}
where $\zeta_L$ is the physical metric field defined with a physical short distance cutoff $\epsilon$ as in \nref{PhysCut}. $G(L)$ is a simple multiplicative constant to ensure that the answer is independent of $L$. With $\langle \rangle_{\text{short}}$ we are similarly integrating out over short-distance modes of $\zeta$, e.g.~the modes between the physical cutoff $\epsilon$ and the fiducial cutoff $\hat \epsilon$ that appears in \nref{Dgeo}. Importantly, this also involves curves $\gamma_p $ which are defined with a physical cutoff $\epsilon$ in \nref{Dprop}, and the coarse-graining will also leave us with the curves $\gamma$ in \nref{Dgeo} that are defined with a coarser fiducial cutoff $\hat \epsilon$. The idea is that this coarse graining should lead to the emergence of the non-trivial critical exponent $\xi$ in \nref{Dgeo}, similarly to what happened for the volume.  This looks like a more complicated computation than the one we did in section \ref{renarg} because it involves taking into account possible spatially dependent correlations of the fast modes, rather than the fast modes at a single $x$ point as in section \ref{renarg}, as well as the minimization over the short distance part of the curve $\gamma_p$. In other words, in \nref{Dprop} the RG is acting on the geometry and also on the paths over which the distance is minimized. 

Let us also mention a related construction in~\cite{Gwynne:2019bhi} based on the dyadic tiling that we reviewed in section \ref{dyadsqr}.  They define a discrete notion of distance using their fundamental dyadic tiles $S$.  If $x$ and $y$ lie in fundamental tiles $S_x$ and $S_y$, respectively, the distance between them is defined as the minimum number of tiles that must be crossed along a path from $S_x$ to $S_y$.  As discussed in section \ref{dyadsqr}, these tiles can be viewed as a discretization of the random geometry into cells of approximately finite physical size.  It is therefore natural to expect that this graph distance provides a discrete approximation to the physical distance appearing in \nref{Dprop}. In fact, \cite{Gwynne:2019bhi} proposed that their construction should be a discrete approximation of \nref{Dgeo}, so we expect the two definitions to match.

\paragraph{Fractal dimension in eternal inflation.} Another historical notion of fractal dimension, which appeared in the context of eternal inflation, is the one given in \cite{Aryal:1987vn} (and refined in \cite{Winitzki:2001np}). The main idea, following \cite{Aryal:1987vn}, is to study the geometry of the universe at a given time slice $\tau$. In eternal inflation, some regions will have reheated already, but some will still be inflating. Therefore, if one considers only the inflating region, the time slice at a given time will seem to have ``reheated'' holes in it. Moreover, the structure of this surface is self-similar under time evolution.

They then interpret the inflating region of the time slices as a fractal, with a non-trivial fractal dimension because of the regions that have reheated. In the context of linear roll inflation, their calculation would go as follows. The probability that inflation is still happening at a time $\tau$ decays exponentially, as $e^{-\frac{dQ^{2}}{4}\tau}$.\footnote{This follows from equation \nref{pstop} in Appendix \ref{statrdmotion}, and the relation of the described random motion with stochastic inflation.} Therefore, since the universe expands exponentially, the volume of the part of the universe still inflating goes as 
\begin{equation}
\label{Vinf}
V_{\text{inf}}\sim e^{d\big(1-\frac{Q^{2}}{4}\big)\tau}\,.
\end{equation}

If we treat each Hubble-sized region as a box, then there are $e^{d\tau}$ boxes on the slice, and we can compute from \nref{Vinf} a box-counting fractal dimension for the inflating region. The fractal dimension $d_{\text{eternal}}$ is then
\begin{equation}
d_{\text{eternal}}=d\bigg(1-\frac{Q^{2}}{4}\bigg).
\end{equation}

It is interesting that in the eternal inflation transition at $Q=2$, $d_{\text{eternal}}$ is exactly zero, meaning that at any given time slice the set where inflation never ends becomes point-like, in the sense that its fractal dimension tends to zero. As we decrease $Q \to 0$, these inflating regions start developing a higher dimension, ultimately approaching the full spatial dimension $d$ so that they completely fill the time slice.

The fractal dimension of \cite{Aryal:1987vn} is interesting from the point of view of eternal inflation, but it seems orthogonal to our discussion of the reheating surface in the main text. For one, the notion of a reheating surface is not very well defined in eternal inflation. Secondly, the fixed time slices they considered have a fixed geometry, so their notion of fractal dimension is really just a measure of how quickly inflating regions grow with time.

\section{A coarse-grained discussion}
\label{discussion}

One lesson of our analysis is that gravity makes the RG time between a physical cutoff and a fiducial cutoff dynamical. The KPZ relation arises from averaging over this fluctuating amount of RG evolution, while the transition to eternal inflation occurs when the corresponding volume-weighted average ceases to exist. In the inflationary setting, this RG time also has a direct interpretation as Lorentzian evolution, making the connection between random geometry and stochastic inflation particularly concrete.

This viewpoint also suggests a way to think about the emergence of geometry. In the gravitational description, the physical metric and a physical short-distance cutoff are fundamental, while the fiducial geometry is an auxiliary structure that becomes useful after coarse-graining. For $Q>2$, integrating out the short-distance fluctuations gives a well-defined continuum description on the smooth fiducial manifold. In this sense, the fiducial geometry emerges as an effective long-distance description of the underlying fluctuating geometry. For $Q<2$, this description breaks down: a fixed physical scale can correspond to an arbitrarily small fiducial scale, and no finite amount of coarse-graining is sufficient to reach it. From this perspective, the transition to eternal inflation can be viewed as a failure of the usual smooth fiducial geometry to emerge from the underlying fluctuating geometry. Conversely, the statement that inflation produces a flat universe can be viewed as the emergence of a flat fiducial geometry as an effective long-distance description.

\subsection*{Acknowledgments}
\indent We thank Leonardo Senatore for detailed explanations and discussions centered on \cite{Dubovsky:2008rf}. 
We also thank Frank Ferrari, Yue-Zhou Li, Nathan Seiberg, Edward Witten,  Matias Zaldarriaga  and ChatGPT for discussions. 

The work of JC is supported by the Simons Collaboration on Celestial Holography, as well as a Fellowship from the Alfred P.~Sloan Foundation.
The work of JM was supported in part by 
U.S. Department of Energy grant DE-SC0009988, and
the Leinweber Forum at the IAS.

\appendix

\section{Statistics over random paths}
\label{statrdmotion}

In section \ref{fokkpl}, we derive the distribution of stopping times of the random motion \nref{rdmotion} described in section \ref{newrenliouv}. An equivalent problem was discussed in \cite{Creminelli:2008es}. We include the derivation here for completeness and to write relevant formulas in our notation. We then use this distribution to compute the relevant expectation values needed in section \ref{newrenliouv}.

In section \ref{markovtrick}, we give a simpler derivation of these expectation values that does not require the explicit stopping time distribution.

\subsection{Fokker-Planck equation and solution}
\label{fokkpl}

Consider the random paths \nref{rdmotion} that started from $N=N_{0}$ at $\tau=0$. The motion stops when it first reaches $N = 0$, so $N = 0$ is an absorbing boundary.  At any time $\tau$, some  branches of random  paths are still evolving with $N> 0$, while the remaining  branches  have already reached the absorbing boundary and stopped.

The sum of the probabilities in both branches is conserved, but the probability in each is not. In fact, as time evolves, more and more paths will have hit $N=0$, so there is a continuous probability flow from the evolving branch into the absorbed branch. Let $P(N,\tau|N_{0})$ be the probability density for the particle to be in position $N$ at time $\tau$ in the evolving branch, given that it started at $N_{0}$ at $\tau=0$. Then, the probability $p(\tau)$ that the particle hits $N=0$ at time $\tau$ is
\begin{equation}
p(\tau)=-\frac{d}{d\tau}\int_{0}^{\infty}dN\,P(N,\tau|N_{0}) 
\end{equation}
where $\int_{0}^{\infty}dN\, P(N,\tau|N_{0})$ is the survival probability, namely the probability that the motion has not yet reached $N = 0$ by time $\tau$.

Therefore, we can solve for $p(\tau)$ by solving for $P(N,\tau|N_{0})$ first. One way to do so is by noting that $P(N,\tau|N_{0})$ satisfies a Fokker-Planck equation.\footnote{From now on, we will sometimes leave the conditioned symbol ``$|N_{0}$'' implicit.} To derive it, consider one infinitesimal time step $\delta \tau$.  If the position changes by $\delta N$, then a path ending at $N$ at time $\tau + \delta \tau$ was at $N - \delta N$ at time $\tau$.  Therefore
\begin{equation}
\begin{gathered}
P(N,\tau+\delta \tau)=\mE[P(N-\delta N,\tau)]=P(N,\tau)+\delta \tau\, \mE\bigg[-\frac{\delta N}{\delta \tau}\partial_{N}P(N,\tau)+\frac{(\delta N)^{2}}{2(\delta \tau)}\partial_{N}^{2}P(N,\tau)\bigg]
\end{gathered}
\end{equation}
where $\mE$ is the average over the paths in the time interval $\delta \tau$. Using $\mE[\delta N] = - \delta \tau$ and $\mE[(\delta N)^2] = \frac{2\,\delta\tau}{d Q^2} + O(\delta \tau^2)$, which follows from \nref{rdmotion}, we obtain
\begin{equation}
\partial_{\tau}P=\partial_{N}P+\frac{1}{dQ^{2}}\partial_{N}^{2}P
\end{equation}
which is the Fokker-Planck equation we wanted to establish.

To solve for $P$, we therefore only need the initial conditions for it and the boundary conditions it must respect. Since the particle started at $N=N_{0}$ at $\tau=0$ we have that
\begin{equation}
P(N,0|N_{0})=\delta(N-N_{0})\,.
\end{equation}
Also, since paths that hit $N=0$ are by definition not in the evolving branch, $P$ should satisfy 
\begin{equation}
P(0,\tau|N_{0})=0
\end{equation}
for any $\tau>0$.  The solution satisfying both conditions can be obtained by the method of images. The resulting solution is then
\begin{equation}
P(N,\tau|N_{0})=\sqrt{\frac{dQ^{2}}{4\pi \tau}}\bigg[\exp\!\bigg(-\frac{dQ^{2}(N-N_{0}+\tau)^{2}}{4\tau}\bigg)-e^{dQ^{2}N_{0}}\exp\!\bigg(-\frac{dQ^{2}(N+N_{0}+\tau)^{2}}{4\tau}\bigg)\bigg].
\end{equation}
The first term is the solution to the problem without the constraint at $N=0$, and the second is an image term. The image term enforces the boundary condition at $N=0$. At $\tau = 0$, the image source is located at $N = - N_0 < 0$, outside the physical domain $N \geq 0$, and therefore does not modify the required initial conditions.

The probability that the path stopped at $N=0$ is therefore
\begin{equation}
\label{pstop}
p(\tau)=-\int_{0}^{\infty} dN\,\partial_{\tau}P=\frac{1}{dQ^{2}}\partial_{N}P|_{N=0}=N_{0}\sqrt{\frac{dQ^{2}}{4\pi \tau^{3}}}e^{-\frac{dQ^{2}(N_{0}-\tau)^{2}}{4\tau}}
\end{equation}
where we used the Fokker-Planck equation, integrated by parts, and imposed the absorbing boundary condition $P(0,\tau | N_0) = 0$. Using this probability distribution, we can derive the expectation value of the operator $e^{d h\tau}$ as
\begin{equation}
\label{expint}
\mE[e^{d h\tau}]=\int d\tau\,p(\tau)e^{d h\tau}=N_{0}\sqrt{\frac{dQ^{2}}{4\pi}}e^{\frac{dQ^{2}N_{0}}{2}}\int_{0}^{\infty} d\tau\,\,\tau^{-\frac{3}{2}}\,e^{-d\big(\frac{Q^{2}}{4}-h\big)\tau-\frac{dQ^{2}N_{0}^{2}}{4\tau}}\,.
\end{equation}
Up to a simple rescaling of $\tau$, the integral can be written as  
\be \la{Aequ}
\int_0^\infty { d \tau \over \tau } \,\tau^{ -1/2} \exp\!\left[ - A (\sqrt{\tau } - 1/\sqrt{\tau } )^2 \right] = \int_{-\infty}^{\infty } dx \, e^{ - A x^2 } ~,~~~~~~~ x =\sqrt{\tau } - 1/\sqrt{\tau  }\,, 
\ee 
where we used the symmetry under $\tau \to 1/\tau$. This implies that 
\begin{equation}
\label{expdeltaAp}
\mE[e^{d h \tau}]= e^{d\alpha_{h} Q N_{0}}
\end{equation}
with
\begin{equation}
\label{alphaap}
\alpha_{h}=\frac{Q}{2}-\sqrt{\frac{Q^{2}}{4}-h}\,,
\end{equation}
where the square root arises from the rescaling of $\tau$.  In particular, in \nref{Aequ} we have $A=\frac{dQN_0}{2}\sqrt{\frac{Q^2}{4}-h}$.

Moreover, we see that $\alpha_{h}$ is the root of
\begin{equation}
h=\alpha_{h}(Q-\alpha_{h})
\end{equation}
that approaches zero as $h\rightarrow0$, which is precisely the KPZ relation. For $h>0$, the moment \nref{expint} is finite precisely when $Q \geq 2\sqrt h$. For $Q<2\sqrt h$, the integral diverges, and the complex value obtained from \nref{alphaap} is simply the analytic continuation of the closed-form expression beyond the domain in which the moment exists. For $h \leq 0$, there is no analogous large-$\tau$ divergence. In section \ref{markovtrick}, we give a simpler derivation of \nref{expdeltaAp}.

\subsection{Expectation values from differential equations}
\label{markovtrick}

In this section, we show how to compute $\mE[e^{d h \tau}]$ using simple points about the random motion \nref{rdmotion}. Since the motion started at $N=N_{0}$, define
\begin{equation}
F[N_{0}]=\mE[e^{d h \tau}]
\end{equation}
where $N_{0}$ stands for the initial condition for the motion. In our context, since there is no special $N$, $N_{0}$ is simply the initial distance to the target, which here is at $N=0$.

We can divide the trajectory into two pieces. Choose $N_1$ with $0<N_1<N_0$, and let $\tau_1$ be the first time the trajectory reaches $N_1$. Let $\tau_2$ be the additional time required to reach $N=0$ after this first crossing of $N_1$. Then the total stopping time is $\tau=\tau_1+\tau_2$.

By the strong Markov property, the motion after the first hitting time of $N_1$ is independent of the earlier trajectory. Moreover, translation invariance in $N$ implies that the first segment is equivalent to a first-passage problem starting a distance $N_0-N_1$ from its target. Therefore,
\begin{equation}
F[N_0] = \mE[e^{dh(\tau_1+\tau_2)}] = F[N_0-N_1]F[N_1]\,.
\end{equation}
The continuous solutions of this multiplicative relation have the form
\begin{equation}
F[N] = e^{\lambda N}.
\end{equation}

To determine the exponent of $N$ in $F$, we use another trick. Say that a motion started at $N$. Then, we can split the motion from $N$ into a first time step of size $\delta \tau$, and a remaining motion of duration $\tau_{\text{new}}$. After the initial time step, the new position of the particle will be $N+\delta N$, so the remaining motion will be a first passage problem that started from this new position. We therefore have that
\begin{equation}
\label{markovtrickF}
F[N]=\mE[e^{d h\delta \tau}F[N+\delta N]]=F[N]+\delta \tau\bigg[d hF[N]+\frac{\mE_{\delta \tau}[\delta N]}{\delta \tau}F'[N]+\frac{1}{2}\frac{\mE_{\delta \tau}[(\delta N)^{2}]}{\delta \tau}F''[N]\bigg]+ \cdots
\end{equation}
where the average $\mE_{\delta \tau}$ is only over the random motion in the small initial time step. Using the random motion equation \nref{rdmotion}, we therefore derive the following differential equation
\begin{equation}
\label{diffeq}
d h\, F-F'+\frac{1}{dQ^{2}}\, F''=0\,.
\end{equation}

Since we know that $F$ is exponential, the differential equation fixes what the exponent is. More conveniently, we can parameterize the exponent as $\lambda= d \alpha_{h} Q$, and so substituting $F[N] = e^{d\alpha_h QN}$ into \nref{diffeq} gives
\begin{equation}
h=\alpha_{h}(Q-\alpha_{h})\,.
\end{equation}

The quadratic equation has two roots. The physical branch is fixed by requiring $\alpha_h \to 0$ as $h \to 0$, so that $F[N] \to 1$ when $h \to 0$. This selects $\alpha_h = \frac{Q}{2} - \sqrt{\frac{Q^2}{4}-h}$, in agreement with \nref{alphaap}. Therefore, we achieved the same solution from a more trivial derivation.

The differential equation \nref{diffeq} also has a direct interpretation in terms of the renormalization of the normal-ordered operator
\begin{equation}
\label{normordap}
\normord{\,e^{d\alpha_h Q\zeta}}\, = e^{d\alpha_h Q\zeta_{\hat\epsilon}} e^{-\frac{d^2\alpha_h^2Q^2}{2}G_{\hat\epsilon}},
\end{equation}
where
\begin{equation}
G_{\hat\epsilon} = \langle\zeta_{\hat\epsilon}^2\rangle = \frac{2}{dQ^2}\log\frac{1}{\hat\epsilon}\,.
\end{equation}
Recall that the gravitational description has an exact redundancy under fiducial Weyl \nref{weylsym}: a change of the fiducial metric can be compensated by a shift of $\zeta$ so that the physical metric remains unchanged. For a constant Weyl transformation, this is equivalently a compensating change of fiducial cutoff $\hat\epsilon\rightarrow e^{-\ell}\hat\epsilon$, $\zeta\rightarrow\zeta+\ell$, which leaves the corresponding physical scale fixed. Under this transformation, $G_{\hat\epsilon}\rightarrow G_{\hat\epsilon}+\frac{2\ell}{dQ^2}$. Therefore,
\begin{equation}
\normord{\,e^{d\alpha_h Q\zeta}} \, \longrightarrow e^{d(\alpha_h Q-\alpha_h^2)\ell} \normord{\,e^{d\alpha_h Q\zeta}}.
\end{equation}
Since the renormalized operator must reproduce the Weyl weight $d\,h$ of the original physical operator $e^{dh\zeta}$, consistency with the exact fiducial-Weyl redundancy requires $d\,h=d\,\alpha_h Q-d\alpha_h^2$, or equivalently $h=\alpha_h(Q-\alpha_h)$, which is again the KPZ relation.

It is instructive to compare this equation directly with the differential equation \nref{diffeq}. Recall from above that substituting $F[N]=e^{d\alpha_h QN}$ into \nref{diffeq} gives $dh-d\alpha_h Q+d\alpha_h^2=0$. The three terms have a simple stochastic interpretation. The term $dh$ comes from the exponential weight $e^{dh\tau}$, the term $-d\alpha_h Q$ comes from the deterministic drift $\mE[\delta N]/\delta\tau=-1$, and the term $d\alpha_h^2$ comes from the Brownian fluctuations $\mE[(\delta N)^2]/\delta\tau=2/(dQ^2)$. Thus the deterministic drift and Gaussian fluctuations in the first-passage problem reproduce, respectively, the classical shift and anomalous normal-ordering contributions to the KPZ scaling dimension.

\bibliographystyle{apsrev4-1long}
\bibliography{main.bib}

\end{document}